\input{epsf}
\documentstyle[eqsecnum,aps,preprint]{revtex}
\tightenlines
\begin{document}

\title{The Quantum Vlasov Equation and its Markov Limit}
\vspace{1.5 true cm}
\author{Yuval Kluger, Emil Mottola \\
Theoretical Division, Los Alamos National Laboratory, \\
MS B285, Los Alamos, New Mexico 87545 {\it USA}\\ and\\
Judah M. Eisenberg\thanks{Deceased}\\
School of Physics and Astronomy \\
Raymond and Beverly Sackler Faculty of Exact Sciences \\
Tel Aviv University, 69978 Tel Aviv, {\it Israel}}
\preprint{LA-UR-98-1138}
\maketitle

\begin{abstract}
\baselineskip 1 pc
The adiabatic particle number in mean field theory obeys a quantum Vlasov equation which is nonlocal in time. For weak, slowly varying electric 
fields this particle number can be identified with the single particle distribution function in phase space, and its time rate of change is the appropriate effective source term for the Boltzmann-Vlasov equation. 
By analyzing the evolution of the particle number we exhibit the time 
structure of the particle creation process in a constant electric field, 
and derive the local form of the source term due to pair creation. 
In order to capture the secular Schwinger creation rate, the source term requires an asymptotic expansion which is uniform in time, and whose longitudinal momentum dependence can be approximated by a delta function 
only on time scales much longer than $\sqrt{p_{\perp}^2 + m^2c^2}/eE$. 
The local Vlasov source term amounts to a kind of Markov limit of field 
theory, where information about quantum phase correlations in the created 
pairs is ignored and a reversible Hamiltonian evolution is replaced by 
an irreversible kinetic one. This replacement has a precise counterpart in 
the density matrix description, where it corresponds to disregarding the rapidly varying off-diagonal terms in the adiabatic number basis and 
treating the more slowly varying diagonal elements as the probabilities 
of creating pairs in a stochastic process. A numerical comparison between 
the quantum and local kinetic approaches to the dynamical backreaction 
problem shows remarkably good agreement, even in quite strong electric
fields, $eE \simeq m^2c^3/\hbar$, over a large range of times.
\thispagestyle{empty}
\end{abstract}
\voffset=1.0 true cm
\newpage

\voffset=-0.5 true cm
\pagestyle{plain}
\pagenumbering{arabic}

\section{Introduction}
\label{sec:level1}

In recent years there has been considerable interest in establishing
the precise connection between quantum field theory and classical kinetic
theory. This interest is motivated by the wide variety of problems in
different fields of physics which require a consistent description
of quantum many body phenomena far from equilibrium. Examples
include chiral symmetry restoration and the quark-gluon plasma phase of QCD, 
soon to be probed by relativistic heavy-ion colliders, baryogenesis
at the electroweak phase transition, and the formation and decay of 
topological defects or Bose condensates, whether in the hot, dense 
early universe, or a cryogenic laboratory environment. 

At their root all these systems may be treated as field theories
with well-defined Hamiltonian evolutions and (except for the
case of explicit CP violation in the electroweak theory) microscopic
time reversal invariance. Yet, a large body of experience confirms the 
macroscopically irreversible behavior of such systems far
from equilibrium, so that it should be possible to approximate the
unitary Hamiltonian evolution of such systems by an irreversible kinetic
description, under suitable circumstances. In addition to the numerous 
potential applications, this raises the fundamental issue of the precise connection between microscopic reversibility and macroscopic
irreversibility which lies at the heart of much of nonequilibrium statistical mechanics. 

Whereas kinetic theory is by far the more developed and familiar framework 
to study nonequilibrium problems, the first steps in their practical solution 
in the context of quantum field theory have been taken only relatively recently
\cite{us1,us2,us3}. Until that had been done it was difficult to even formulate the question of the relationship between the field theory and kinetic theory approaches in a clear way. As a practical matter the kinetic description is certainly the simpler one to formulate and implement numerically on a computer. However, the Boltzmann-Vlasov equation essentially describes classical point particles, and extensions to quantum collective phenomena, time-evolving mean 
fields and off-shell virtual processes, which are quite natural in field 
theory, present considerable difficulties for a purely kinetic approach. 
Also lost in the kinetic description from the very outset is a detailed
understanding of how time reversible Hamiltonian evolution comes to be 
replaced by time irreversible dissipative behavior. For these reasons of 
both fundamental interest and practical application, our purpose in this 
paper is to explore the precise relationship between the two approaches in a concrete example. 

In the interest of being as clear and specific as possible we focus our attention in this paper on charged particle creation in electric 
fields, a phenomenon which was discussed nearly seventy years ago by 
Klein and Sauter, and twenty years later by Schwinger as a prime example 
of the then newly developed theory of quantum electrodynamics \cite{Sauter}. Over the years there has developed an extensive literature on the topic \cite{NarNik,BreItz,MarPop,PerMos}, which has continued to attract interest up to the present time \cite{GavGit,CasNeu,GleMat,BiaCzy,BBGR,Best,Rau,RauMueller,Rost}. Several monographs summarizing this activity have also appeared \cite{books}. Given this background it might be 
supposed that no aspect of particle creation in electric fields 
has been left uncovered. The reason that this is not quite true
is that interest in the real time evolution of particle creation
and its incorporation into transport theory by an effective source
term is relatively recent. 

In many treatments of particle 
creation, analytic continuation of the amplitudes to complex time have been employed. Though elegant and useful in other contexts, complex continuation methods cannot address directly the real time evolution 
of the particle creation event and thus cast little light on the source term for a kinetic description. A suggestion of how to incorporate the Schwinger pair creation mechanism in the context of kinetic theory was first made in 1979, based on an intuitively appealing picture of the instantaneous semiclassical creation event \cite{CasNeu}. This mechanism has been a subject of renewed interest in the context of heavy ion collisions and QCD due to the suggestion that the receding ions might produce a strong chromoelectric flux tube between them which shorts itself out by the creation quark/anti-quark pairs (see {\it e.g.} \cite{us2} and references therein). The ansatz of ref. \cite{CasNeu} has been taken over to the QCD flux tube model as well.
Yet it should be clear from the outset that a delta function source term which requires that the charged particles be created at precisely zero momentum, 
at a definite instant of time can only be an approximation to the rapid 
but continuous evolution of wave amplitudes in the underlying quantum theory.
Calculations of the backreaction of the charged particle 
pairs on the electric field in QED in a well defined continuous evolution
were compared with the {\it ad hoc} kinetic theory, according to the ansatz of ref. \cite{CasNeu}. Reasonable qualitative agreement between the mean field evolutions in the two approaches was found, although they certainly differ in 
quantitative detail, such as in the distributions of created particles \cite{us2}. In these numerical investigations the time structure of the individual creation events was not addressed, leaving open the question of
the limit of validity of the delta function ansatz for the source term.

The Wigner function formalism has also been proposed \cite{deGroot,Heinz} 
as a method for deriving relativistic transport equations from the 
underlying field theory. It has become increasingly 
clear however, that the covariant Wigner function does not readily 
lend itself to practical calculation, because covariance requires splitting
the time variable in the Wigner transformation in parallel to the
splitting of the spatial variable, with the consequence that the problem
ceases to be well posed as an evolution from initial data.
More recently, an alternative, noncovariant formalism, in which the time variable is not split has been suggested \cite{BBGR,Best}.
As has been emphasized in earlier work the lack of manifest covariance
is not a problem since the initial value description 
of even a relativistic field theory in Hamiltonian terms is necessarily 
noncovariant in form, but the evolution equations are
completely equivalent to those derived from a covariant action
principle \cite{us3}. In any case, a firm conclusion about the
source term has not been obtained by these investigations either. 
Finally, the general projection formalism of Zwanzig \cite{Zwa} has been advocated as a route to a transport description of particle creation \cite{Rau,RauMueller,Rost}, although the time structure of the creation process itself has not been investigated in detail in this approach, and the conditions of validity of the delta function approximation for a local source term in the Vlasov equation has remained obscure.
  
By revisiting the electrodynamic pair creation problem our purpose in
this paper is to elucidate fully the precise connection between the field theoretic and kinetic treatments in this particular case. 
Application and extension of our methods of incorporating particle 
creation into a kinetic description for the other situations of interest 
will then become possible. Our first step will be to specify 
completely the adiabatic particle number basis in which particle creation 
can be described as a phase interference (or {\em dephasing}) phenomenon 
of the quantum theory from the effective Hamiltonian point of view \cite{us3}. Writing the explicit Bogoliubov transformation to this adiabatic particle
basis then identifies a time dependent particle number whose
total change recaptures the Schwinger formula in a constant,
uniform electric field, and whose time derivative yields the 
appropriate source term for the Boltzmann-Vlasov equation. 
The adiabatic particle number obeys a nonlocal quantum Vlasov
equation, and in this sense is completely consistent with the
general approach advocated in refs. \cite{Rau,RauMueller,Rost}. 
The relationship of our method to that of the projection formalism
may be seen most clearly by considering the density matrix
in the adiabatic particle number basis. However
we have no need for the general projection formalism, since
the source term for the Vlasov equation can be written in closed form 
in terms of the wave functions of the charged particle modes in the background constant electric field. In this way we derive for the first time a local 
form for the source term, which explicitly exhibits the relationship to the semiclassical picture of particles spontaneously appearing out of the 
vacuum in real time. The electromagnetic current of the charged particle 
pairs also has a simple form in this basis, corresponding to a clear
physical interpretation in terms of a quasiclassical conduction
current and the quantum polarization current of particle creation.
The fact that the current grows linearly in time for a fixed external
electric field and that therefore backreaction must eventually
become important even for arbitrarily small coupling is also easy
to see in the adiabatic particle basis. This will also serve to clarify
the nature of the ``time divergences" discussed in refs. \cite{GavGit}.

The essential physical ingredient in passing from the quantum unitary 
evolution to the irreversible Vlasov description is the dephasing phenomenon,
{\it i.e.} the near exact cancellation of the rapidly varying
phases of the quantum mode functions contributing to the mean electric 
current of the created pairs. This cancellation depends in turn upon a clean separation of the time scales, 
\begin{enumerate}
\item $\tau_{qu}$, of the very rapidly oscillating modes of the microscopic 
quantum theory,
\item $\tau_{cl}$, of the more slowly varying mean number of particles in the adiabatic number basis, and
\item $\tau_{pl}$, of the collective plasma oscillations of the electric 
current and mean electric field produced by those particles. 
\end{enumerate}
In the limit $\tau_{qu} \ll \tau_{cl}$ quantum coherence 
between the created pairs can be neglected because of efficient dephasing 
and a (semi)classical local kinetic approximation to the underlying quantum
theory becomes possible. In the limit $\tau_{cl} \ll \tau_{pl}$
the electric field may be treated as approximately
{\it constant} over the interval of particle creation. Thus
when both inequalities apply we can replace
the true nonlocal source term which describes particle creation
in field theory by one that depends only on the instantaneous
value of the quasistationary electric field, at least over very
long intervals of time.  

The essential mathematical ingredient in the exploitation of this hierarchy 
of time scales is an asymptotic expansion of the wave functions and particle number for constant electric fields 
uniformly valid on the real time axis, so that secular particle creation
effects (which are lost in the usual nonuniform WKB expansion)
are retained. It is this precise sense of evaluating
the effect of rapid degrees of freedom on slow degrees of 
freedom by treating the latter as constant in leading order of
a uniform asymptotic expansion (which recalls the Born-Oppenheimer 
approximation in atomic and molecular physics) and deriving a local
effective source term for the change of adiabatic particle number
that we refer to as the Markov limit of the quantum Vlasov equation.
 
The importance of a {\it uniform} asymptotic expansion of the
wave functions is that secular particle creation effects are
retained in an expansion valid everywhere on the real time
axis. The true wave functions exhibit a sharp change
in amplitude, on the time scale $\tau_{cl}$, at or near the time 
of the semiclassical creation event which is captured very well by 
a uniform asymptotic expansion in terms of Airy functions. 
As we shall see, if one is interested only in the collective phenemena 
on time scales of $\tau_{pl}$ or longer, then the details of the particle creation process on the time scale $\tau_{cl}$ are unimportant and one can replace the momentum distribution of the source term by one localized
at zero kinetic momentum, as has been the practice in the earlier phenomenological approaches, provided only that the integrated
distribution gives the correct total creation rate. This
will clarify the precise conditions of validity of such an\"atze for
the first time.   
 
Since an asymptotic (not a convergent) expansion is involved, the limit 
of the ratio of time scales $\tau_{qu}/\tau_{cl} \rightarrow 0$ for
fixed $t$ and the long time limit $t \rightarrow \infty$ of the evolution 
for fixed ratio $\tau_{qu}/\tau_{cl}$ do {\it not} commute in general. 
Hence for any small but finite ratio $\tau_{qu}/\tau_{cl}$ there can 
be eventually a very large but finite $t$ at which the quantum phases reassemble and the irreversible local kinetic description breaks down. 
Up to this very long (typically exponentially and possibly infinite) recurrence time the system behaves in many practical respects like an irreversible one, in which the quantum phase coherence between the created pairs appears to have been lost. In this way the apparent incongruity of an effectively irreversible time evolution emerging from a unitary Hamiltonian field theory is removed.

The paper is organized as follows. In the next section we review
(scalar) QED mean field theory in the leading order
of the large $N$ expansion. By exhibiting explicitly the Hamiltonian
structure of these equations we demonstrate that they are completely
time reversible. In Section 3 we define
the adiabatic particle number basis which is selected by the
Hamiltonian evolution and derive the exact nonlocal form of the 
quantum Vlasov equation for this quantity. The quantum density
matrix in this basis is also derived. In Section 4 we solve for the 
source term of the Vlasov equation in the limit of constant mean electric 
field, and study the pair creation process for this case in some detail.
It is shown that particle creation in a fixed external field produces
an electric current which grows linearly with time, so that any
amount of particle creation (no matter how small) eventually requires
a substantial backreaction on the field in any self-consistent treatment.
In Section 5 the technique of uniform asymptotic expansions 
for the mode functions and adiabatic particle number is brought to bear. 
The source term for particle creation in a constant field is calculated 
to leading order in this asymptotic expansion in terms of Airy functions 
and yields an effectively Markovian source term for the {\it local} Vlasov equation describing pair creation in weak, slowly varying electric fields. 
The circumstances under which further approximation of the Airy function 
source term by an instantaneous delta function source term becomes 
permissable is also discussed. In Section 6 the dynamical backreaction 
problem for the charged particles whose current is self-consistently 
coupled to the mean electric field is compared to the two (Airy and 
delta function) local approximations for the Vlasov source term in the 
kinetic description, and relatively good agreement is obtained. 
We close with a summary of our results and some concluding remarks on 
possible generalizations of the analysis to other systems of interest. 
The derivation of the density matrix in the adiabatic particle number 
basis is relegated to an Appendix.
   
\section{Scalar QED in the Large $N$ Limit}
\label{sec:level2}

Let us begin by reviewing the equations of motion for scalar
QED in a uniform electric field in the semiclassical limit in which
the matter field is fully quantized and the electromagnetic field
is treated classically. This limit can be obtained in a consistent way
by taking the leading order of a large $N$ expansion (where $N$ is
the number of identical copies of the charged matter field) \cite{us1,us2}. 
We take the electric field spatially homogeneous, and express the 
vector potential in the gauge,
\begin{equation}
{\bf A} = A(t) {\bf \hat z}\,,\qquad {\rm A_0 = 0}\,,
\end{equation}
so that the electric field is
\begin{equation}
{\bf E} = - \dot A {\bf \hat z} = E {\bf \hat z}\,.
\end{equation}
The charged scalar field operator is expanded in Fourier modes
in Fock space in the usual way,
\begin{equation}
\Phi ({\bf x}, t) = {1\over \sqrt{V}} \sum_{\bf k} e^{i\bf k \cdot x}  
\varphi_{\bf k}(t) = {1\over \sqrt{V}} \sum_{\bf k} \left\{ 
e^{i\bf k \cdot x} f_{\bf k}(t) a_{\bf k} 
+ e^{-i\bf k \cdot x} f^*_{-\bf k}(t) b^{\dagger}_{\bf k}\right\}\ . 
\end{equation}
The time-independent creation and destruction operators obey
the commutation relations
\begin{equation}
[a_{\bf k}, a^{\dagger}_{\bf k'}] = [b_{\bf k}, b^{\dagger}_{\bf k'}] 
= \delta_{\bf k, k'}\,
\label{aadag}
\end{equation}
in the finite large volume $V$, and the Fourier components,
\begin{equation}
\varphi_{\bf k}(t) \equiv f_{\bf k}(t) a_{\bf k} + 
f^*_{\bf k}(t) b^{\dagger}_{-\bf k}
\label{fcoor}
\end{equation} 
may be regarded as (complex) generalized coordinates of the field $\Phi$ 
for the purposes of the Hamiltonian description. The momentum canonically
conjugate to this coordinate is
\begin{equation}
\pi_{\bf k}(t) = \dot\varphi_{\bf k}^{\dagger}(t) = 
\dot f^*_{\bf k}(t) a_{\bf k}^{\dagger} + 
\dot f_{\bf k}(t) b_{-\bf k}\,,
\label{fmom}
\end{equation} 
which obeys the canonical commutation relation,
\begin{equation}
[\varphi_{\bf k}, \pi_{\bf k'}] = i\hbar \delta_{{\bf k}, {\bf k'}}\,,
\end{equation} 
provided that the mode functions satisfy the Wronskian condition,
\begin{equation}
f_{\bf k} \dot f^*_{\bf k} - \dot f_{\bf k} f^*_{\bf k}  = i \hbar\,,
\label{Wron}
\end{equation}
and (\ref{aadag}) is used.  

The time dependence in this basis is carried by the
complex mode functions $f_{\bf k}(t)$ which satisfy the equations of motion,
\begin{equation}
\left({d^2\over dt^2} + \omega^2_{\bf k}(t)\right) f_{\bf k}(t) = 0
\label{modeq}
\end{equation}
where the time dependent frequency $\omega^2_{\bf k}(t)$ is given by
\begin{equation}
\omega^2_{\bf k}(t) = \left( {\bf k} -e {\bf A}\right)^2 + m^2 =
(k-e A(t))^2 + k_{\perp}^2 + m^2\,.
\end{equation}
Here $k$ is the constant canonical momentum in the $\bf\hat z$ direction
which should be clearly distinguished from the gauge-invariant 
but time dependent {\em kinetic} momentum,
\begin{equation}
p(t) =k-eA(t)\quad ; \qquad \dot p = -e \dot A = e E
\label{cankin}
\end{equation}
which reflects the acceleration of the charged
particle due to the electric field. In the directions transverse to
the electric field the kinetic and canonical momenta are the same
and do not need to be distinguished, {\em i.e.} we shall
use the notation $p_{\perp} = k_{\perp}$ interchangeably.
When expressed as a function of the kinetic momenta we use the
notation, $\omega(p, p_{\perp}) = \sqrt{p^2 + p_{\perp}^2 + m^2}$,
or simply $\omega$.

The mean value of electromagnetic current in the $\bf\hat z$ direction is 
\begin{equation}
j(t) = 2e \int\, [d{\bf k}] \left(k-eA(t)\right) 
\vert f_{\bf k}(t)\vert^2 (1 + N_{+}({\bf k}) + N_{-}(-{\bf k}))
\label{curr}
\end{equation}
where
\begin{eqnarray}
N_{+}({\bf k}) &\equiv& \langle a^{\dagger}_{\bf k}a_{\bf k}\rangle\nonumber\\
N_{-}({\bf k}) &\equiv& \langle b^{\dagger}_{\bf k}b_{\bf k}\rangle
\label{Hpart}
\end{eqnarray}
are the mean numbers of particles and antiparticles in the time independent
basis and
\begin{equation}
{1\over V} \sum_{\bf k} \rightarrow \int [d{\bf k}] \equiv \int {d^3{\bf k}\over
(2\pi)^3}
\end{equation}
in the infinite volume continuum limit. We make use of the freedom
in defining the initial phases of the mode functions to set the
correlation densities $\langle a_{\bf k}a_{\bf k}\rangle =
\langle b_{\bf k}b_{\bf k}\rangle = 0$, without any loss of generality.

The mean charge density must vanish,
\begin{equation}
j^0(t) = e \int\, [d{\bf k}] \left[N_{+}({\bf k}) - 
N_{-}(-{\bf k})\right] = 0\,, 
\label{char}
\end{equation}
by Gauss' Law for a spatially homogeneous electric
field ($i.e. {\bf \nabla \cdot E} = 0$). We shall further restrict ourselves 
to the subspace of states for which 
\begin{equation}
N_{+}({\bf k}) = N_{-}(-{\bf k}) \equiv N_{\bf k}
\label{special}
\end{equation}
for simplicity in what follows, although this is a stronger condition than 
is required by Eq. (\ref{char}). Clearly the vacuum $N_{+}({\bf k}) = 
N_{-}(-{\bf k}) =0$ (as well as a thermal mixed state) belongs to this 
class of states.

Self-consistent evolution of the mean electric field requires coupling
it to the expectation value of the current of the charged field by the only nontrivial Maxwell equation remaining in this homogeneous example, namely,
\begin{equation}
-\dot E = \ddot A = j = 2e \int\, [d{\bf k}] \left(k-eA(t)\right) 
\vert f_{\bf k}(t)\vert^2 (1 + 2N_{\bf k} )\, .
\label{max}
\end{equation}
For the analysis of the source term in a constant electric field
and its uniform expansion in the next three sections we will
treat the electric field as fixed and nondynamical, returning to
Eqn. (\ref{max}) and the dynamical backreaction problem in Section VI.
  
By a slight change of notation it is possible to recast the mean field
evolution equations (\ref{modeq}) and (\ref{max}) together with the
quantum Wronskian condition (\ref{Wron}) as Hamilton's equation for an effective classical Hamiltonian in which $\hbar$ appears as a parameter. Defining the real quantities,
\begin{eqnarray}
\sigma_{\bf k} &\equiv& 1 + N_{+}({\bf k}) + N_{-}(-{\bf k}) = 1 + 2N_{\bf k}\,,\nonumber\\
\xi^2_{\bf k}(t) &\equiv& \sigma_{\bf k}\vert f_{\bf k}(t)\vert^2\,,
\qquad {\rm and}\nonumber\\
\eta_{\bf k}(t) &\equiv & \dot \xi_{\bf k}(t)
\label{xietadef}
\end{eqnarray}
we find that the mode equation (\ref{modeq}) can be rewritten in the
form,
\begin{equation}
\ddot \xi_{\bf k} = \dot{\eta}_{\bf k} = - \omega_{\bf k}^2 \xi_{\bf k} +
{\hbar^2 \sigma_{\bf k}^2 \over 4 \xi_{\bf k}^3}\, ,
\label{xieom}
\end{equation}
when account is taken of (\ref{Wron}). This last equation together with
the Maxwell equation (\ref{max}) will be recognized as Hamilton's
equations for the Hamiltonian,
\begin{equation}
H_{eff} (A, p_A; \{\xi_{\bf k}\}, \{\eta_{\bf k}\} ; \{\sigma_{\bf k}\}) =
{V\over 2}E^2 + \sum_{\bf k} \left( \eta_{\bf k}^2 + 
\omega_{\bf k}^2\xi_{\bf k}^2 + {\hbar^2 \sigma_{\bf k}^2 \over 4\xi_{\bf k}^2}
\right)\,,
\label{heff}
\end{equation}
where $p_A \equiv -E$ is the momentum conjugate to $A$ and $\eta_{\bf k}$
is the momentum conjugate to $\xi_{\bf k}$. 

Moreover, the quantum statistical density matrix of the charged scalar field
corresponding to the mean field evolution can be written as a product of Gaussians in Fourier space, {\it viz.}
\begin{equation}
\langle \{\varphi_{\bf k}^{\prime}\}|{\bf\rho} |\{\varphi_{\bf k}\}\rangle =
\prod_{\bf k }\langle\{\varphi_{\bf k}^{\prime}\}\vert{\bf\rho}
(\xi_{\bf k}, \eta_{\bf k}; \sigma_{\bf k})\vert\{\varphi_{\bf k}\}\rangle 
\equiv \prod_{\bf k }\rho_{\bf k}
\end{equation}
with
\begin{equation}
\rho_{\bf k} =(2\pi \xi_{\bf k}^2)^{-{1\over 2}}
\exp \biggl\{ -{\sigma_{\bf k}^2 + 1\over 4\xi_{\bf k}^2}\left[ \varphi_{\bf
k}^{\prime\ast}\varphi^{\prime}_{\bf k} + \varphi_{\bf k}^{\ast}\varphi_{\bf k}\right] +i\,{\eta_{\bf k} \over \hbar\xi_{\bf k}}\left[\varphi_{\bf k}^{\prime\ast}\varphi_{\bf k}^{\prime}- \varphi_{\bf k}^{\ast}\varphi_{\bf k}\right]  + {\sigma_{\bf k}^2 - 1\over 4 \xi_{\bf k}^2} \left[\varphi_{\bf k}^{\prime\ast}\varphi_{\bf k} + \varphi_{\bf k}^{\prime} \varphi_{\bf k}^{\ast}\right] \biggr\}~,
\label{gaussd} 
\end{equation}
and $\varphi_{\bf k}$ is the complex generalized coordinate of the classical field in Fourier space, defined by (\ref{fcoor}) (with $a_{\bf k}$
and $b_{-\bf k}$ treated as c-numbers). The Liouville equation for the evolution of this density matrix according to the quantum
Hamiltonian of a free charged scalar field in a background electric potential, 
\begin{equation}
\dot \rho = -i[H_{qu},\rho]~ ; \qquad H_{qu} = {1\over 2}\sum_{\bf k} \left( \pi_{\bf k} \pi_{\bf k}^{\dagger}+ 
\omega_{\bf k}^2\varphi_{\bf k}\varphi_{\bf k}^{\dagger} + h.c. \right)
\label{Liov}
\end{equation}
gives precisely the equations of motion (\ref{xieom}) for the width parameters
of the time dependent Gaussian. The effective classical Hamiltonian 
(\ref{heff}) is nothing else 
than the expectation value of the quantum Hamiltonian of scalar QED $H_{qu}$ 
in the Gaussian density matrix $\rho$, {\em i.e.} $H_{eff} = 
{\rm Tr} (\rho H_{qu})$. Notice that
in this Schr\"odinger representation of the time evolution all
the equations are local in time, {\it i.e.}, they involve a single time argument, and there is no need to introduce Wigner functions with two time arguments, although these correlation functions at unequal times may be calculated easily enough from knowledge of the density matrix, if desired.
In contrast to several earlier approaches to kinetic theory from
field theory principles \cite{deGroot,Heinz}, we shall not require these unequal time correlators or Wigner functions.

The constant parameters $\sigma_{\bf k} = 1 + 2N_{\bf k}\ge 1$ measure the
extent that the quantum state is a mixed state. If $N_{\bf k} = 0$,
$\sigma_{\bf k} = 1$ and the state is pure, as is evident from the
vanishing of the last term in (\ref{gaussd}), so that the
density matrix becomes a simple product $\vert \psi\rangle\langle \psi\vert$.
In either the pure or more general mixed state the density matrix
(\ref{gaussd}) possesses a $U(1)$ symmetry under
\begin{eqnarray}
\varphi_{\bf k} &\rightarrow & \varphi_{\bf k}\, \exp({i \zeta_{\bf k}})
\nonumber\\
\varphi_{\bf k}^{\prime} &\rightarrow &\varphi_{\bf k}^{\prime} 
\,\exp ({i \zeta_{\bf k}})
\end{eqnarray}
for each $\bf k$. This is a reflection of the fact that the generator
of the local $U(1)$ gauge transformation of electrodynamics for a
spatially uniform electric field is the charge density (\ref{char}),
and we have restricted ourselves to charge symmetric states obeying 
(\ref{special}), so that the density matrix has this $U(1)$ invariance
in each Fourier mode independently.
 
In this leading order of the large $N$ expansion the density matrix of 
the electric field is also a Gaussian and multiplies the matter field Gaussian above, so that the evolution of the closed system with the backreaction Eqn. (\ref{max}) is also Hamiltonian. Clearly the Hamiltonian evolution
equations (\ref{modeq}), with or without the Maxwell Eqn. (\ref{max}) are
completely time reversible upon reversing the signs of all the momenta.

Forgetting for the moment the Maxwell equation of backreaction on the
electric field we see that the mean field evolution is 
equivalent to a set of time-dependent harmonic oscillators, with
a different time-dependent frequency $\omega_{\bf k}(t)$ for each
Fourier mode $f_{\bf k}$. Treating these frequencies as arbitrary,
slowly varying functions of time we may write down the
Hamilton-Jacobi equation corresponding to the effective classical
Hamiltonian $H_{eff}$, namely,
\begin{equation}
\left({d W_{\bf k} \over d \xi_{\bf k}}\right)^2 + \omega_{\bf k}^2 
\xi_{\bf k}^2 + {\hbar^2 \sigma_{\bf k}^2 \over 4 \xi_{\bf k}^2} = 
\epsilon_{\bf k}\,,
\end{equation}
and find that the Hamilton principal function $W_{\bf k}$ evaluated over
one full period,
\begin{equation}
{W_{\bf k} \over 2 \pi \hbar} = 
{1\over 2 \pi \hbar}\oint d\xi_{\bf k} \sqrt{2\epsilon_{\bf k} - 
\omega_{\bf k}^2\xi_{\bf k}^2 - 
{\hbar^2\sigma_{\bf k}^2\over 4\xi_{\bf k}^2}} = {\epsilon_{\bf k} \over 2\hbar \omega_{\bf k}} 
- {\sigma_{\bf k}\over 2} 
\end{equation}
is an adiabatic invariant of the periodic motion. Since $\sigma_{\bf k}$
is strictly a constant for all $\bf k$, this implies that
\begin{equation}
{\epsilon_{\bf k}(t)\over \hbar\omega_{\bf k}(t)} \equiv 2{\cal N}_{\bf k} (t) 
+ 1
\label{adbinv}
\end{equation}
is an adiabatic invariant of the motion for slowly varying $\omega_{\bf k}(t)$.
It is this adiabatic invariant that defines a time-dependent particle number 
basis which becomes the appropriate one for making contact
with the Boltzmann-Vlasov kinetic description of particle creation.

\section{The Adiabatic Number Basis}
\label{sec:level3}

From the field theory development of the last section we note that
$N_+, N_-$ and $k$, appear quite naturally in either the time independent (Heisenberg) or time dependent (Schr\"odinger) descriptions as constants
of motion under the Hamiltonian evolution. 
However, kinetic theory is expressed in terms
of time-evolving quantities ${\cal N}_+ (t), {\cal N}_- (t)$
and $p(t)$ which must be clearly distinguished from the analogous time
independent quantities above. The difference between the canonical and 
kinetic momenta $k$ and $p(t)$ in (\ref{cankin}) is clear enough on basic kinematic grounds. The specification of the time dependent particle numbers ${\cal N}_+ (t)$ and ${\cal N}_- (t)$ may not be quite as obvious, but as they 
provide the essential connection between the field theory and kinetic descriptions we must take special care to be equally clear and explicit about their definition. This requires that we introduce a Bogoliubov transformation 
from the time independent to a time-dependent (but adiabatic) number basis.

The observation underlying the introduction of this basis is that
the mode equation (\ref{modeq}) generally posseses time dependent
solutions which have no clear {\em a priori} physical meaning in
terms of particles or antiparticles. The familiar
notion that positive energy solutions to the wave equation
correspond to particles while negative energy solutions correspond
to antiparticles is quite meaningless in time dependent background fields
where the energy of individual particle/antiparticle modes is not
conserved, and no such neat invariant separation into positive
and negative energy solutions of the wave equation is possible. This 
is just a reflection of the fact that physical particle number does not 
correspond to a sharp operator which commutes with the Hamiltonian, {\em i.e.}
particle/antiparticle pairs are created or destroyed, and physical 
particle number is not conserved in time dependent background fields. 
  
Given this fact, one possible point of view is to forget completely
about particle number in time-dependent backgrounds and deal only with 
conserved physical currents like $j(t)$ in (\ref{curr}). Indeed, in 
arbitrarily strong and rapidly time-varying fields this is the only 
possible point of view, since all notion of even an approximately conserved particle number disappears, and there is no possibility
whatsoever of a classical kinetic description in such extreme situations.
One must rely then exclusively on the field theoretic framework.

When the fields are not quite so strong and/or so rapidly varying in time
we would expect to be able to define a particle number which varies slowly enough for the comparison to an effective semiclassical kinetic description 
to be meaningful. Clearly this physical slowly varying particle number is
{\it not} the $N_{\bf k}$ of the time independent Heisenberg basis defined
by (\ref{Hpart}) above, since this $N_{\bf k}$ is part of the initial
data, a strict constant of the equations of motion, no matter how strong or rapidly varying the electric field is. The physical particle number
at time $t$ must be defined instead with respect to a
time dependent basis that provides some criterion to distinguish particles from antiparticles at the instant $t$. This time dependent basis (the adiabatic number basis) permits a semiclassical correspondence limit to ordinary positive
energy plane wave solutions in the limit of slowly varying 
$\omega_{\bf k}(t)$, and is related to the Heisenberg basis by a time dependent 
Bogoliubov transformation. Since the mode equation (\ref{modeq}) 
is the equation of motion of a (complex) harmonic oscillator with time varying
frequency $\omega_{\bf k}(t)$, governed by the effective classical
Hamiltonian $H_{eff}$ of (\ref{heff}) standard arguments from classical Hamilton-Jacobi theory inform us that there is an adiabatic invariant proportional to the energy of the oscillator divided by its frequency,
and given by (\ref{adbinv}). It is this quantity which can be used to define 
an adiabatic particle number and to make the connection with classical kinetic theory. Corresponding to this slowly varying action variable there is a conjugate angle variable which is rapidly varying, of which classical kinetic theory takes no account.  

The adiabatic basis is defined by first constructing the adiabatic
mode functions,
\begin{equation}
\tilde f_{\bf k}(t) \equiv \sqrt {\hbar\over 2 \omega_{\bf k}(t)}
\exp\left(-i\int^t \omega_{\bf k}(t') dt'\right)\,.
\label{adbmod}
\end{equation}
We will make use of the shorthand notation for the phase,
\begin{equation}
\Theta_{\bf k}(t) \equiv \int^t \omega_{\bf k}(t') dt'\ ,
\label{actvar}
\end{equation}
suppressing the explicit dependence on $t$ (and occasionally also the
momentum index $\bf k$) except when needed for clarity
in most of the following. The lower limit of the integral in (\ref{actvar}) 
and therefore also the absolute phase of the mode function $\tilde f$ are
left arbitary for the moment, to be fixed in a convenient way
in the next section. It is clear that in the limit of arbitrarily weak 
electric fields $\omega_{\bf k}(t)$ becomes independent of time 
and can be removed from the integral in (\ref{actvar}). In that limit the
adiabatic mode function becomes the usual positive energy plane
wave solution with respect to which the usual definition of
particle number is taken. Otherwise the adiabatic mode functions (\ref{adbmod})
will not be exact solutions of the mode equation (\ref{modeq}), but
we are still free to specify a basis with respect to them, provided only
that $\omega_{\bf k}(t)$ remains real and positive for all $\bf k$ and $t$. 

The transformation to this basis from the original one
is specified by the two linear relations,
\begin{eqnarray}
f_{\bf k}(t) &=& \alpha_{\bf k}(t) \tilde f_{\bf k}(t) + \beta_{\bf k}(t)
\tilde f^*_{\bf k}(t)\nonumber\\
\dot f_{\bf k}(t) &=& -i\omega_{\bf k}\alpha_{\bf k}(t) \tilde f_{\bf k}(t) 
+ i\omega_{\bf k}\beta_{\bf k}(t)\tilde f^*_{\bf k}(t)
\label{Bog}
\end{eqnarray}
between the exact and adiabatic mode functions. When the phase
of $\tilde f$ is fixed these relations completely fix the complex coefficients $\alpha_{\bf k}(t)$ and $\beta_{\bf k}(t)$. It is straightforward to solve for the Bogoliubov coefficients directly in the form,
\begin{eqnarray}
\alpha_{\bf k} &=& i (\dot f_{\bf k} - i \omega_{\bf k} f_{\bf k}) 
\tilde f^*_{\bf k}\qquad {\rm and}\nonumber\\
\beta_{\bf k} &=& -i (\dot f_{\bf k} + i \omega_{\bf k} f_{\bf k}) 
\tilde f_{\bf k}\,.
\label{alpbet}
\end{eqnarray}
An equivalent form of this Bogoliubov transformation in the Fock space of creation and destruction operators is
\begin{eqnarray}
a_{\bf k} &=& \alpha_{\bf k}^*(t) \tilde a_{\bf k}(t) - \beta^{*}_{\bf k}(t)
\tilde{b}^{\dagger}_{-\bf k}(t)\nonumber\\
b_{-\bf k}^{\dagger} &=& \alpha_{\bf k}(t) \tilde{b}^{\dagger}_{-\bf k}(t) - \beta_{\bf k}(t) \tilde a_{\bf k}(t)\,,
\label{bog}
\end{eqnarray}
so that the field coordinate $\varphi_{\bf k}(t)$ may be expressed
equally well in the time independent basis by (\ref{fcoor}) or
in the time dependent basis by
\begin{equation}
\varphi_{\bf k}(t) = \tilde f_{\bf k}(t) \tilde a_{\bf k}(t) 
+ \tilde f^*_{\bf k}(t) \tilde b^{\dagger}_{-\bf k}(t)\,,
\label{fcooradb}
\end{equation}
and likewise the field momentum variable is given either by (\ref{fmom}) or by
\begin{equation}
\pi_{\bf k}(t) = -i\omega_{\bf k}(t)
\tilde f_{\bf k}^*(t) \tilde a_{\bf k}^{\dagger}(t) 
+ i\omega_{\bf k}(t) \tilde f_{\bf k}(t) \tilde b_{-\bf k}(t)\,.
\label{fmomadb}
\end{equation}
The transformation from the time independent
$(a_{\bf k}, b^{\dagger}_{-\bf k})$ basis to the time dependent
adiabatic basis $(\tilde a_{\bf k}, \tilde b^{\dagger}_{-\bf k})$ 
requires two independent relations (\ref{Bog}) or (\ref{bog}),
corresponding to a canonical transformation in a two dimensional 
(complex) phase space, for which
\begin{equation}
\vert\alpha_{\bf k}\vert^2 - \vert\beta_{\bf k}\vert^2 = 1
\label{can}
\end{equation}
for each $\bf k$. It is easily verified that (\ref{alpbet}) satisfies this
relation when the Wronskian condition (\ref{Wron}) is used. Because of
(\ref{can}) the magnitude of the Bogoliubov transformation to
the adiabatic number basis $\gamma_{\bf k}(t)$ may be specified by
\begin{eqnarray}
\vert\alpha_{\bf k}(t)\vert &=& {\rm cosh} \gamma_{\bf k}(t)\,,\nonumber\\
\vert\beta_{\bf k}(t)\vert &=& {\rm sinh} \gamma_{\bf k}(t)\,.
\label{magbog}
\end{eqnarray}

We now define the adiabatic particle number to be
\begin{eqnarray}
{\cal N}_{\bf k}(t) &\equiv& \langle \tilde a_{\bf k}^{\dagger} (t) \tilde 
a_{\bf k}(t)\rangle \nonumber\\  
&=& \vert\alpha_{\bf k}\vert^2 \langle a^{\dagger}_{\bf k} a_{\bf k}\rangle
+ \vert\beta_{\bf k}\vert^2 \langle b_{-\bf k} b^{\dagger}_{-\bf k} \rangle
\nonumber\\
&=&\left( 1 + \vert\beta_{\bf k}\vert^2\right) N_+({\bf k}) + \vert\beta_{\bf k}\vert^2 \left( 1 + N_-(-{\bf k})\right)\nonumber\\
&=& N_{\bf k} + \left( 1 + 2N_{\bf k}\right)\,\vert\beta_{\bf k}(t)\vert^2
\nonumber\\
&=& N_{\bf k} + \left( 1 + 2N_{\bf k}\right)\,{\rm sinh}^2 \gamma_{\bf k}(t)\,.
\label{adbpar}
\end{eqnarray}
The second of the relations (\ref{Bog}) is essential to define the
adiabatic basis in which particle number is given by the ratio
of energy to frequency. In fact, 
\begin{eqnarray}
{\epsilon_{\bf k}(t)\over\hbar\omega_{\bf k}(t)} &=& 
\left(1 + 2 N_{\bf k} \right)
{\left(\vert \dot f_{\bf k}\vert^2 + \omega_{\bf k}^2 \vert f_{\bf k}\vert^2\right)\over\hbar\omega_{\bf k}}\,\nonumber\\
&=&\left( 1 + 2N_{\bf k}\right) \left( 1 + 2\vert\beta_{\bf k}\vert^2\right)\,
\nonumber\\
&=& 1 + 2{\cal N}_{\bf k}(t) \,.
\label{enerom}
\end{eqnarray}
 
Hence the particle number ${\cal N}_{\bf k}(t)$, though time dependent, 
is an adiabatic invariant of the motion. 
Consequently, it is the natural candidate for a particle density in 
phase space for a kinetic description, becoming the ordinary asymptotic constant particle number in the limit of slowly varying $\omega_{\bf k}(t)$. 
Although this choice of basis is not unique, since we could have chosen 
a different condition on $\dot f_{\bf k}$ in (\ref{Bog}),
it is the only basis where the ratio $\varepsilon_{\bf k}/ \omega_{\bf k}$
is simply related to particle number (without the appearance of
$\dot\omega_{\bf k}$ or higher derivative terms, for example) 
which is the standard adiabatic invariant of the harmonic 
oscillator with time dependent frequency as in (\ref{adbinv}). 
In different contexts (such as particle creation in external gravitational fields) it may be appropriate to consider slightly different definitions of
the adiabatic number basis.

Now that we have completely specified the time dependent particle
number basis it is straightforward to derive the equation
of motion which it obeys. We note that from the explicit representation
(\ref{alpbet}) by differentiation and use of the mode equation (\ref{modeq})
we have
\begin{eqnarray}
\dot \alpha_{\bf k} &=& {\dot \omega_{\bf k}\over 2 \omega_{\bf k}} \beta_{\bf k}
\exp (2i\Theta_{\bf k})\,,\qquad {\rm and}\nonumber\\
\dot \beta_{\bf k} &=& {\dot \omega_{\bf k}\over 2 \omega_{\bf k}} \alpha_{\bf k}\exp (-2i\Theta_{\bf k})\,.
\label{abeom}
\end{eqnarray}
These two first order differential equations are entirely equivalent
to the second order mode equation in Hamiltonian form. We now obtain
by differentiating (\ref{adbpar}),
\begin{eqnarray}
{d\over dt}{\cal N}_{\bf k} &=& 2\left( 1 + 2N_{\bf k}\right){\rm Re}\,
(\beta_{\bf k}^*\dot\beta_{\bf k})\nonumber\\
&=& {\dot\omega_{\bf k}\over\omega_{\bf k}}\left( 1 + 2N_{\bf k}\right)\,{\rm Re}
\left\{\alpha_{\bf k}\beta_{\bf k}^*\exp (-2i\Theta_{\bf k})\right\}\nonumber\\
&=& {\dot\omega_{\bf k}\over\omega_{\bf k}}\, {\rm Re} \left\{{\cal C}_{\bf k}
\exp (-2i\Theta_{\bf k})\right\}\,,
\label{Neom}
\end{eqnarray}
where we have defined the time dependent pair correlation function,
\begin{eqnarray}
{\cal C}_{\bf k}(t) &\equiv& \langle \tilde a_{\bf k}(t) \tilde b_{\bf -k}(t)\rangle
\nonumber\\
&=& \left( 1 + 2N_{\bf k}\right)\alpha_{\bf k} \beta_{\bf k}^*\,.
\label{pcor}
\end{eqnarray}
Thus the time derivative of the adiabatically slowly varying particle
number involves the pair correlation function ${\cal C}_{\bf k}(t)$ which is 
itself very rapidly varying, since the time dependent phases on the right side of (\ref{pcor}) {\em add} rather than cancel, although the phases do nearly
cancel in the final combination of (\ref{Neom}). The time derivative of
the pair correlation function, 
\begin{eqnarray}
{d\over dt} {\cal C}_{\bf k} &=& {\dot\omega_{\bf k}\over 2\omega_{\bf k}} 
\left( 1 + 2N_{\bf k}\right) \exp (2i\Theta_{\bf k})\,\left( 1 + 2\vert\beta_{\bf k}\vert^2\right)\nonumber\\
&=&{\dot\omega_{\bf k}\over 2\omega_{\bf k}} 
\left( 1 + 2{\cal N}_{\bf k}\right) \exp (2i\Theta_{\bf k})\,,
\end{eqnarray}
brings us back again to ${\cal N}_{\bf k}$.
This last equation may be solved formally for ${\cal C}_{\bf k}$ and
subsituted into (\ref{Neom}) to obtain
\begin{equation}
{d\over dt}{\cal N}_{\bf k} = {\dot\omega_{\bf k}\over 2\omega_{\bf k}} 
\int^t_{t_0}\,dt'\,\left\{ {\dot\omega_{\bf k}\over \omega_{\bf k}}(t') 
\left( 1 + 2{\cal N}_{\bf k}(t')\right) \cos \left[2\Theta_{\bf k}(t) 
-2\Theta_{\bf k}(t')\right]\right\}\,,
\label{nonl}
\end{equation} 
where we have assumed that ${\cal C}_{\bf k}$ vanishes at some $t=t_0$
(which could be taken to $-\infty$).

Equation (\ref{nonl}) may be be called a ``quantum Vlasov equation,"
in the sense that it gives the quantum creation rate of particle number 
in an arbitary time varying mean field. Let us remark that the Bose 
enhancement factor $( 1 + 2{\cal N}_{\bf k})$ appears in (\ref{nonl}), 
so that both spontaneous and induced particle creation are included 
automatically in the quantum treatment. The most important feature
of Eqn. (\ref{nonl}) for our present purpose is that it is nonlocal
in time, the particle creation rate depending on the entire previous history
of the system. In that sense the particle creation process is
certainly non-Markovian in general \cite{Rau,Rost}.
Eqn. (\ref{nonl}) becomes exact in the limit in which the electric field 
can be treated classically, {\em i.e.} the large $N$ limit in which real and virtual photon emission is neglected, and there is no scattering.  
Inclusion of scattering processes lead to collision
terms on the right side of (\ref{nonl}) which are also nonlocal in general. 
This nonlocality is essential to the quantum description in which 
phase information is retained for all times.
The phase oscillations in the cosine term are a result of the 
quantum coherence between the created pairs, which must be
present in principle in any unitary evolution. However, precisely
because these phase oscillations are so rapid it is clear that the 
integral in (\ref{nonl}) receives most of
its contribution from $t'$ close to $t$, which suggests that some
local approximation to the integral should be possible, provided that we
are not interested in resolving the short time structure or
measuring the phase coherence effects. The time scale for these
quantum phase coherence effects to wash out is the time scale of several
oscillations of the phase factor $\Theta_{\bf k}(t) - \Theta_{\bf k}(t')$,
which is of order $\tau_{qu}={2\pi/\omega_{\bf k}} = {2\pi\hbar/\epsilon_{\bf k}}$,
where $\epsilon_{\bf k}$ is the single particle energy. 

The steps we have just performed to arrive
at (\ref{nonl}) are a special case of the general
projection formalism of Zwanzig \cite{Zwa}, where some subset of fast dynamical variables 
deemed ``irrelevant" (in this case ${\cal C}_{\bf k}$) are eliminated in favor 
of slow variables deemed ``relevant" (in this case ${\cal N}_{\bf k}$). 
Because the two variables are coupled by the underlying Hamiltonian equations of motion the result of solving for some variables in terms of others
is generally nonlocal in time. The nonlocal form (\ref{nonl})
is still completely equivalent to the mode equation (\ref{modeq})
and absolutely nothing has been lost (or gained) by this rewriting.
In other words, the projection method is essentially free of any physical
content, until and unless one makes further approximations that replace
the nonlocal relations satisfied by the relevant observables by
{\em local} ones. It is at this point that great care must be exercised,
since the precise form of the local approximation made will determine
the usefulness and range of validity of the resulting truncation.

For example, one's first idea might be to remove the Bose enhancement
factor $1 + 2{\cal N}_{\bf k}(t')$ from the integral, or simply
ignore it entirely, on the basis that it is slowly varying function 
(for real $t'$), and attempt to perform the remaining integral over the rapidly varying phase by the method of stationary phase. However, this does not lead to a useful result, since the phase becomes stationary at $\dot \Theta_{\bf k} = \omega_{\bf k} = 0$ which is precisely where the integrand has a pole in the complex $t'$ plane. This is more than just an inconvenience since the vanishing of $\omega_{\bf k}$ is the condition of a turning point, where the adiabatic approximation certainly breaks down. Hence it is in 
just the neighborhood of this point in the complex 
$t'$ plane where the dominant contribution to the integral arises that the 
removal of the factor $1 + 2{\cal N}_{\bf k}(t')$ from the integrand or
its neglect cannot be justified. The stationary phase method has been carried out nonetheless for certain background field problems where integrals similar to that in (\ref{nonl}) arise, with the result that the correct exponential
factor but the {\em in}correct prefactor is obtained \cite{BreItz}. 

The importance of the complex turning point(s) for determing the asymptotic
mixing between particle and antiparticle modes as $t \rightarrow \pm \infty$
has been emphasized by Marinov and Popov in refs. \cite{MarPop}. 
In their method the analytic continuation of the solutions of the mode 
equation around the Stokes' lines emanating from the turning point in the complex time plane determines the subdominant component of the wave function with the opposite sign of the frequency on the real axis. The amplitude of
this exponentially subdominant component of antiparticle waves
in the wavefunction is the Schwinger particle creation effect. 
However, the method outlined by  these authors does not seem to be 
applicable to the integral in (\ref{nonl}) directly, since it is designed 
for calculating the particle creation asymptotically over infinite time, 
not for determining the evolution of the particle creation process in finite real time $t$, which is what we require for the transport description. 

Finally, if one takes no account of the stationary phase point in the
complex $t'$ plane but attempts to approximate the integral in (\ref{nonl})
entirely in real time, for example by integrating the rapidly varying
cosine function by parts any number of times, it is easy to see that an asymptotic series is generated in which the exponentially small subdominant solution can {\em never} appear after any finite number of such steps.
Any asymptotic expansion of the wavefunction on the real axis which
discards the exponentially small antiparticle component will miss
the Schwinger creation effect at late times.
      
From this discussion we see that the essential difficulty with
Eqn. (\ref{nonl}) is that the point(s) in the complex $t'$ plane where the
phase $\Theta_{\bf k}$ is stationary must play the critical role in
determining the particle creation for asymptotically late times, 
but we cannot evaluate the contribution to the integral of these stationary 
phase points where $\omega_{\bf k}$ vanishes without in effect 
knowing the full ${\cal N}_{\bf k}, \omega_{\bf k}$ and $\Theta_{\bf k}$
as analytic functions in the entire complex $t$ plane
before we even begin. If we were in possession of these
analytic functions we would already have the full solution to
our dynamical problem, without any need to make any approximations to
the integral. This is clearly impossible except for a small number of special
cases where the complete analytic structure is known {\em a priori}.
Thus the nonlocal form of the quantum Vlasov equation (\ref{nonl})
makes it difficult to extract any useful information about a
source term for a kinetic description in general.

Consideration of this difficulty immediately suggests a different approach. 
Instead of trying to work with the nonlocal equation (\ref{nonl}), in the
next section we evaluate the spontaneous pair creation rate ${d\over dt} 
{\cal N}_{\bf k}(t)$ for a {\it constant} electric field 
analytically and directly in real time, thereby assuring agreement with the Schwinger result in both its exponential and nonexponential factors. This
is one of special cases where ${\cal N}_{\bf k}$ and its time
derivative can be evaluated analytically in local form, directly from the definition (\ref{adbpar}) without any need for the nonlocal integral representation (\ref{nonl}). Then by making use of an asymptotic expansion of the exact analytic result for constant fields, uniformly valid everywhere on the real time axis, we obtain a useful {\em local} approximation to the spontaneous pair creation rate for the slowly varying electric fields, 
without any need for analytic continuation or stationary phase methods in complex time. By such an approach we shall bypass completely the difficulties of dealing with the nonlocal integral equation (\ref{nonl}) resulting from the projection method.

The transformation to the adiabatic number basis and elimination of 
the rapid variables ${\cal C}_{\bf k}$ in favor of the slow variables
${\cal N}_{\bf k}$ by Eqns. (\ref{pcor})-(\ref{nonl}) has its counterpart 
in the density matrix description as well. It is shown in the 
Appendix that the density matrix (\ref{gaussd}) may be transformed to
the adiabatic number basis, with the general form of the nonvanishing
matrix elements given by (\ref{rhopart}). In the pure state case
$\sigma_{\bf k} = 1$ the only nonvanishing matrix elements of
$\rho$ are in uncharged pair states with equal numbers of positive
and negative charges, $\ell_{\bf k} = n_{\bf k}^{(+)} = n_{\bf k}^{(-)}$, with $\ell_{\bf k}$ the number of pairs in the mode $\bf k$, {\it viz.}
\begin{equation}
\langle 2\ell'_{\bf k} \vert \rho \vert 2\ell_{\bf k}\rangle \bigg\vert_{_{\sigma =1}} =
e^{i({\ell_{\bf k}}^{\prime}- \ell_{\bf k})\vartheta_{\bf k} (t) }\ {\rm sech}^2 \gamma_{\bf k} (t)\ \left({\rm tanh}\gamma_{\bf k} (t)\right)^{{\ell_{\bf k}}^{\prime} + \ell_{\bf k}} 
\label{adbden}
\end{equation}
where the magnitude of the Bogoliubov transformation, $\gamma_{\bf k}(t)$ is defined by (\ref{magbog}) and its phase, $\vartheta_{\bf k} (t)$ is specified by
\begin{equation}
\alpha_{\bf k}\beta_{\bf k}^{\ast} e^{-2i\Theta_{\bf k}}
= - {\rm sinh}\gamma_{\bf k}\ {\rm cosh}\gamma_{\bf k}\ e^{i\vartheta_{\bf k}}\,.
\label{thetdef}
\end{equation}
Hence the off-diagonal matrix elements $\ell' \neq \ell$ of $\rho$ are
rapidly varying on the time scale $\tau_{qu}$ of the
quantum mode functions, while the diagonal matrix
elements $\ell' = \ell$ depend only on the adiabatic 
invariant average particle number via 
\begin{equation}
\langle 2\ell_{\bf k} \vert \rho \vert 2\ell_{\bf k}\rangle 
\bigg\vert_{_{\sigma =1}} \equiv
\rho_{2\ell_{\bf k}} ={\rm sech}^2 \gamma_{\bf k} {\rm tanh}^{2\ell_{\bf k}} \gamma_{\bf k} = {\vert\beta_{\bf k}\vert^{2\ell_{\bf k}}
\over (1 + \vert\beta_{\bf k}\vert^2)^{\ell_{\bf k}+1}} = {{\cal N}_{\bf k}^{\ell_{\bf k}}\over
(1 + {\cal N}_{\bf k})^{\ell_{\bf k}+1}}\bigg\vert_{_{\sigma =1}}\,,
\label{rhodiag}
\end{equation}
and are therefore much more slowly varying functions of time. 
The average number of positively charged particles (or negatively
charged antiparticles) in this basis is given of course by 
\begin{equation}
\sum_{\ell_{\bf k}=0}^{\infty}\ell_{\bf k} \rho_{2\ell_{\bf k}} 
= {\cal N}_{\bf k}\,.
\end{equation}
Thus the diagonal and off-diagonal elements of the density matrix
in the adiabatic particle number basis stand in precisely the
same relationship to each other and contain the same information
as the particle number ${\cal N}_{\bf k}$
and pair correlation ${\cal C}_{\bf k}$ respectively. 

Using the representation (\ref{adbden}) or (\ref{rhodiag}) we can
understand how entropy can increase and the evolution become
time irreversible if we replace the exact nonlocal quantum Vlasov
equation (\ref{nonl}) by a local expression in which the rapid
phase variables ${\cal C}_{\bf k}$, $\vartheta_{\bf k}$, or the
off-diagonal matrix elements of $\rho$ no longer appear.
Time reversal in the field theory requires that both the
slow and fast variables be time reversed, which involves the
full density matrix $\rho$. If we restrict attention to only
the diagonal matrix elements of $\rho$ in the adiabatic particle
number basis without any account of the phase information
present in the rapidly varying off-diagonal elements, then
time reversal no longer holds. In the effective density matrix 
(\ref{rhodiag}) the diagonal elements $\rho_{2\ell_{\bf k}}$ may be 
interpreted (for $\sigma_{\bf k} = 1$) as the independent
probabilities of creating $\ell_{\bf k}$ pairs of charged particles 
with canonical momentum $\bf k$ from the vacuum. This corresponds to
disregarding the intricate quantum phase correlations between the created
pairs in the unitary Hamiltonian evolution, and treating the
creation events as essentially independent in a stochastic Markovian 
processes. Thus the Markov approximation to the field theory arises 
quite naturally when the quantum density matrix is expressed in the 
adiabatic particle number basis.

Such an approximation is known to be quite accurate for long
intervals of time in the backreaction of the current on the 
electric field producing the pairs, for the simple reason that
the phase information in the pair correlations cancels very efficiently
when one considers the sum over all the $\bf k$ modes in the current
(\ref{curr}). It is for this reason that for practical purposes one can approximate the full Gaussian density matrix over large time intervals 
by its diagonal elements only, in this basis. Naturally this truncation
of the unitary Hamiltonian evolution according to (\ref{Liov})
leads to a nonunitary irreversible evolution in which the {\it effective} 
von Neumann entropy of the diagonal density matrix (\ref{rhodiag}),
\begin{equation}
S_{eff}(t) = -{\rm Tr}\,\rho_{eff} \ln \rho_{eff} = -\sum_{\bf k}\sum_{\ell_{\bf k} = 0}^{\infty} \rho_{2\ell_{\bf k}} \ln \rho_{2\ell_{\bf k}}
\end{equation}
can increase with time. In fact, upon subsituting (\ref{rhodiag}),
the sums over $\ell_{\bf k}$ are geometric series which are easily performed,
with the result that the von Neumann entropy of this truncated
density matrix,
\begin{equation}
S_{eff}(t)\Big\vert_{_{\sigma =1}} = \sum_{\bf k} \left\{
(1 + {\cal N}_{\bf k})\ln (1 + {\cal N}_{\bf k}) - 
{\cal N}_{\bf k}\ln {\cal N}_{\bf k}\right\} 
\end{equation}
is precisely equal to the Boltzmann entropy of the single particle
distribution function ${\cal N}_{\bf k}(t)$. Hence
\begin{equation}
{d \over dt}S_{eff} = \sum_{\bf k} \ln \left( {1 + {\cal N}_{\bf k}
\over{\cal N}_{\bf k}}\right) {d \over dt}{\cal N}_{\bf k}
\label{entropy}
\end{equation}
increases if the mean particle number increases. This is always
the case {\it on average} if one starts with vacuum initial conditions, $\sigma_{\bf k} = 1$, since $\vert \beta_{\bf k}\vert^2$ is
necessarily nonnegative and can only increase if it is zero
initially \cite{Kan}. Locally, or once particles are present in the initial state, there is no reason why particle number or the entropy (\ref{entropy}) must continue to increase monotonically in time, and indeed
small temporary decreases are observed in backreaction simulations \cite{us3}. Hence there is no Boltzmann H-theorem for the
effective entropy (\ref{entropy}) without introducing some explicit time averaging and/or further assumptions into the scheme. 
  
Before closing this section we wish to take note of
one additional especially simple property of the adiabatic particle 
number basis. Inserting the Bogoliubov transformation of the mode functions 
(\ref{Bog}) into the expression for the current (\ref{curr}) we obtain
\begin{eqnarray}
j(t) = e \int\, [d{\bf k}] {\left(k-eA(t)\right)\over \omega_{\bf k}(t)} 
(1+ 2\vert \beta_{\bf k}(t)\vert^2 + 
2{\rm Re}\{\alpha_{\bf k}\beta^{\ast}_{\bf k} e^{-2i\Theta_{\bf k}(t)}\} )
(1 + 2N_{\bf k}) \, .
\label{curr2}
\end{eqnarray}
We note that the vacuum term in this expression,  
$\int\, [d{\bf k}] {\left(k-eA(t)\right)\over \omega_{\bf k}(t)}$ vanishes
by charge conjugation symmetry,
when proper gauge invariant integration boundaries are chosen.
Using the mean value of particles in the adiabatic number basis
(\ref{adbpar}), its time derivative and the equations of motion
(\ref{Neom}), we can rewrite the current (\ref{curr2}) as
\begin{eqnarray}
j(t) &=& 2e \int\, [d{\bf k}] {\left(k-eA(t)\right)\over \omega_{\bf k}(t)} 
{\cal N}_{\bf k}(t)
+{2 \over E} \int\, [d{\bf k}] \omega_{\bf k}(t) \dot{{\cal N}}_{\bf k}(t)
\nonumber\\
&=& j_{cond} + j_{pol} \, .
\label{curr3}
\end{eqnarray}
On the other hand, from a classical point of view if the particle distribution ${\cal N}_{\bf k}$ is coupled to a uniform electric field
the energy density and its time derivative
are given by
\begin{eqnarray}
\varepsilon &=& {E^2\over 2}+ 2 \int\, [d{\bf k}] \omega_{\bf k} 
{\cal N}_{\bf k}
\label{epsilon}
\\
\dot{\varepsilon} &=& \dot{E}E+
+ 2 \int\, [d{\bf k}] \left(eE{(k -eA)\over\omega_{\bf k}}{\cal N}_{\bf k} + \omega_{\bf k} \dot{\cal N}_{\bf k}\right)=0\, .
\label{epsilondot}
\end{eqnarray}
Using the Maxwell equation $-\dot{E}=j$ this last relation is precisely
the {\em same} as the mean value of the quantum current in
(\ref{curr3}). Hence we may identify the
adiabatic particle number ${\cal N}_{\bf k}(t)$ with the 
(quasi)classical single particle distribution. Other definitions
of time varying particle number, such as that used in our own earlier
work \cite{us1} do not have this property or admit this simple
quasiclassical interpretation. This exercise also demonstrates that 
the two terms in the mean current (\ref{curr3}) should indeed be interpreted 
as the conduction and polarization terms of the earlier phenomenological
descriptions.
 
\section{Constant Electric Field}
\label{sec:level4}

In order to derive the source term due to particle creation in a
slowly varying electric field, we first analyze the time structure
of the creation process in a constant, uniform electric field, for which
\begin{equation}
A(t) = - E t.
\end{equation} 
It is useful to define the rescaled dimensionless variables
\begin{equation}
u \equiv \epsilon {k +eEt\over \sqrt{\vert eE\vert}} = {\epsilon p(t)\over \sqrt{\vert eE\vert}} \qquad {\rm and} \qquad \lambda \equiv
{k_{\perp}^2 + m^2 \over \vert eE \vert} > 0\,,
\label{rescaled}
\end{equation}
where $\epsilon = \epsilon (eE) = \pm 1$ is the sign of $eE$.
Then the mode equation (\ref{modeq}) may be put into the
form,
\begin{equation}
\left({d^2 \over du^2} + u^2 + \lambda\right) f = 0
\label{waveq}
\end{equation}
whose solutions are parabolic cylinder (Weber) functions.
In fact, the two complex conjugate pairs of solutions,
\begin{eqnarray}
f_{(+)}(u) = f_{(-)}^*(u) &\propto & D_{-{1\over 2} + i{\lambda\over 2}}
\left(-(1-i)u\right)\qquad {\rm and}\nonumber \\
f^{(+)}(u) = f^{(-)*}(u) &\propto & D_{-{1\over 2} - i{\lambda\over 2}}
\left((1+i)u\right)
\label{parabolic}
\end{eqnarray}
each comprise complete sets of basis functions in which to expand
the scalar charged field $\Phi$. Normalizing these solutions
according to the Wronskian condition (\ref{Wron}) and
defining the phase,
\begin{equation}
\psi \equiv {\lambda\over 4} - {\lambda\over 4} \ln \lambda
+{\lambda\over 4} \ln 2 - {\pi \over 8}
\end{equation}
we can write the properly normalized positive frequency mode functions
in the form,
\begin{eqnarray}
f_{(+)\bf k}(t) &=& \vert 2eE\vert^{-{1\over 4}} e^{-{\pi \lambda\over 8}}
e^{i\psi}\, D_{-{1\over 2} + i{\lambda\over 2}}
\left(-(1-i)u\right)\nonumber \quad {\rm and}\\
f^{(+)}_{\bf k}(t) &=& \vert 2eE\vert^{-{1\over 4}} e^{-{\pi \lambda\over 8}}
e^{-i\psi}\, D_{-{1\over 2} - i{\lambda\over 2}}
\left((1+i)u\right)
\label{exactmode}
\end{eqnarray}
which approach the adiabatic functions $\tilde f_{\bf k} (t)$ in the asymptotic limits 
$t\rightarrow -\infty$ and $t \rightarrow \infty$ respectively. 
Notice that with $u$ defined including the $\epsilon$ function as in 
(\ref{rescaled}) these limits are equivalent to $u \rightarrow -\infty$ and $u \rightarrow \infty$ respectively, independently of the sign of $eE$.
The complex conjugates of these solutions are the corresponding
negative frequency mode functions and are denoted by $f_{(-)\bf k}$
or $f_{\bf k}^{(-)}$ respectively. The phase $\psi$ has been defined 
in such a way that the phase of the exact mode functions $f_{\bf k}$
agrees with the adiabatic mode functions (\ref{adbmod}) 
with phase $\Theta_{\bf k}$ 
measured from the symmetric point $u = 0$, {\it i.e.}
\begin{eqnarray}
\Theta_{\bf k} (t) &=& \int _{u=0}^t dt'\omega_{\bf k} (t') = \int_0^u du' 
{\sqrt{u'^2 + \lambda}}\nonumber\\
&=& {1\over 2} u {\sqrt{u^2 + \lambda}} + {\lambda\over 2} \ln
\left( {u + {\sqrt{u^2 + \lambda}}\over \sqrt\lambda}\right)\,. 
\label{Tk}
\end{eqnarray}

It may seem surprising at first sight that the exact mode functions
approach the adiabatic ones in the infinite past and infinite future
even though the electric field $E$ is constant and never vanishes
in these limits. The reason for this is that the corrections
to the lowest-order adiabatic mode functions involve 
\begin{equation}
{\delta \omega_{\bf k}^2(t) \over \omega_{\bf k}^2(t)}
= {1\over 2} {\ddot \omega_{\bf k}\over \omega_{\bf k}^3} - {3\over 4}
{\dot \omega^2_{\bf k}\over \omega_{\bf k}^4} =
{(-3u^2 + 2\lambda) \over 4 (u^2 +\lambda)^3}
\end{equation}
which goes to zero like $|t|^{-4}$ as $t \rightarrow \pm \infty$.

If the state of the system is the vacuum {\it in} state then the
mode function $f_{\bf k}$ to be used in (\ref{alpbet}) is the
$f_{(+) \bf k}$ of (\ref{exactmode}) and the effective source
term for the creation of particles from the vacuum is
\begin{eqnarray}
{d \over dt}{\cal N}_{\bf k}\bigg\vert_{N_{\bf k} =0}=  
{d \over dt}\vert{\beta}_{\bf k}\vert^2  
= \vert 8eE\vert^{-{1\over 2}} e^{-{\pi \lambda\over 4}}
\, {\partial\over \partial t}\left\{{1\over \omega_{\bf k}}\ 
\Bigg\vert \left({\partial\over \partial t} + i\omega_{\bf k}
\right)D_{-{1\over 2} + i{\lambda\over 2}}\left(-(1-i)u\right)\Bigg\vert^2\right\}\,.
\label{newsource}
\end{eqnarray}
We note that for a strictly constant electric field this is
an {\it exact} result for the rate of adiabatic particle number
change starting from vacuum initial conditions at $t=-\infty$.
Phase correlation information for this particular initial state
has not been discarded, although the pair correlation function does not 
appear explicitly in (\ref{newsource}) which is {\it local} in time. 

Now since the two pairs of complex functions $f_{(\pm)\bf k}$
and $f_{\bf k}^{(\pm)}$ both satisfy the same second order
wave equation there exist linear relations between them. Indeed
it follows from the properties of the Weber parabolic cylinder 
functions that
\begin{equation}
f_{(+)\bf k} = \bar \alpha f_{\bf k}^{(+)} + \bar \beta f_{\bf k}^{(-)}
\label{connection}
\end{equation}
with \cite{GraRyz,Erd} 
\begin{eqnarray}
\bar \alpha &=& {\sqrt {2\pi}\over \Gamma \left({1-i\lambda\over 2}\right)}
e^{2i\psi + {i\pi \over 4}} e^{-{\pi\lambda\over 4}}\qquad {\rm and}\nonumber\\
\bar\beta &=& -i e^{-{\pi\lambda\over 2}}
\label{betabar}
\end{eqnarray}
The fact that $\bar\beta \ne 0$ is the statement that the Bogoliubov
transformation between the two basis pairs is nontrivial and
the adiabatic vacuum state in the infinite past contains particle--antiparticle
pairs with respect to the adiabatic vacuum state in
the infinite future. The magnitude of this total Bogoliubov transformation
from $t = -\infty$ to $t =+\infty$ is finite and given by
\begin{equation}
\vert\bar\beta\vert^2 \equiv \sinh^2 \bar\gamma = e^{-\pi \lambda},
\label{schw}
\end{equation}
which is independent of $k$ in the direction of the electric field.

By transforming the Gaussian density matrix corresponding to the
evolution of the charged scalar field in a background electric
field one can show that (\ref{schw}) is also the
mean number of particles in the final state with
respect to the {\it out} vacuum, assuming that the field was
prepared in the {\it in} vacuum. The details of this transformation
are given in the Appendix. From the result, (\ref{rhodiag}) with
$\gamma_{\bf k}$ replaced by $\bar\gamma$ and the discussion of
the previous section discarding the rapidly varying off-diagonal
elements of $\rho$, we may interpret the diagonal elements as
the probability of finding $\ell$ pairs at late times if none were present 
initially. Hence the $\ell =0$
matrix element,
\begin{equation}
{\rm sech}^2 \bar \gamma = (1 + e^{-\pi\lambda})^{-1}
\end{equation}
is the probability of creating no pairs in the given mode, and the
probability that the vacuum remains the vacuum in the future is
given by the product over all modes,
\begin{equation}
\prod_{\bf k}(1 + e^{-\pi\lambda})^{-1} 
= \exp\left( -\sum_{\bf k} \ln (1 + e^{-\pi\lambda})\right)\,.
\end{equation}
Taking the infinite volume limit this can be expressed as $\exp (-VT\Gamma)$
where the rate of vacuum decay per unit volume is
\begin{equation}
\Gamma = {1\over T} \int [d{\bf k}] \ln (1 + e^{-\pi\lambda})\,.
\label{rate}
\end{equation}
Since the kinetic momentum of the created charged particles
in the direction of the electric field is $k + eEt$, 
the longitudinal integration element $dk$ can be replaced by $eET$ 
as $T \rightarrow\infty$ and the vacuum decay rate becomes
\begin{eqnarray}
\Gamma &=& {eE\over (2\pi)^3} \int d^2 {\bf k}_{\perp} 
\ln (1 + e^{-\pi\lambda})
\nonumber\\
&=& {eE\over (2\pi)^3} \int d^2 {\bf k}_{\perp} \sum_{n=1}^{\infty}
{(-)^n\over n} e^{-\pi n \lambda}\nonumber\\
&=& {(eE)^2\over (2\pi)^3}\sum_{n=1}^{\infty}
{(-)^{n+1}\over n^2} e^{-{\pi n m^2\over \hbar eE}}
\label{Srate}
\end{eqnarray}
which is Schwinger's result for scalar QED. 

One should note that the replacement of the longitudinal 
momentum integral over $k$ in (\ref{rate}) by $eET$ in the large $T$ 
limit can be justified only if one understands the time evolution of the
pair creation event, for otherwise the expression (\ref{rate})
is formally meaningless. This replacement of $\int dk$ by $eET$
and the resulting finite expression (\ref{Srate}), which can be
obtained by quite different methods imply that only those $k$ in a 
linearly growing window in time actually contribute to the rate, although 
the mixing coefficient $\bar\beta$ over all time is independent of $k$.
It is the time dependent evolution of $\beta_{\bf k} (t)$
which we can investigate in detail with our definition of the time 
dependent adiabatic number basis in the next section. This
definition smoothly interpolates between the {\it in} and {\it out} vacuum states specified respectively by the two wave functions in (\ref{exactmode}),
so that $\beta_{\bf k} (t)$ starts at zero as $t \rightarrow -\infty$ and
must approaches $\bar\beta$ as $t \rightarrow +\infty$. 
The wave functions depend on $k$ and $t$ only through the variable
$u$ defined in (\ref{rescaled}), and the potential $u^2 + \lambda$ is even
in $u$. Hence we should expect each $k$ mode to go through its
creation event at a different time $t$ according to $k +eEt \approx 0$,
{\it i.e.} for the particles to be created with kinetic momenta near zero.
We shall see that this is indeed the case and that therefore the
range of $k$ which have gone through the creation process at time
$t$ depends linearly on $t$, which justifies the passage from
(\ref{rate}) to (\ref{Srate}).

Omitting the integration over ${\bf k}_{\perp}$ and the phase space 
factor ${1/ (2\pi)^3}$ in (\ref{Srate}), 
one obtains the probability per unit time per 
unit volume to produce pairs with transverse momentum ${\bf k}_{\perp}$
\cite{NarNik,CasNeu}. This result has been interpreted 
as the rate at which pairs are created \cite{GleMat}
and used as a source term in Vlasov equation, which involves particle 
production \cite{BiaCzy}. However, a necessary condition for this interpretation to be correct is that the time integration
over the rate of particle production
\begin{equation}
eE \int_{-\infty}^{\infty} dt \ln (1 + e^{-\pi\lambda})
\label{intprob}
\end{equation}
be identical to the total number of particles produced
per unit volume with transverse momentum ${\bf k}_{\perp}$,
which is given by integration over $k$ of Eq. (\ref{betabar})
\begin{equation}
\int\, dk\, e^{-\pi\lambda}= 
eE\int\, dt\, e^{-\pi\lambda}\,.
\label{intrate}
\end{equation}
Expressions (\ref{intprob}) and (\ref{intrate}) are {\it not} equivalent, because the probability rate of particle production differs from
the production rate of the mean value of particles. They become
equal only in the limit of large $\lambda$ when both the
probability and mean number of produced particles become very small.
It is clear that the source term for the mean rate of particle production
in the Vlasov equation should involve the latter quantity (\ref{intrate}) in principle, without the appearance of any logarithm in the final answer.

\section{Uniform Asymptotic Expansion of the Source Term}
\label{sec:level5}

Eqn. (\ref{newsource}) is the source term due to particle creation,
with a specific choice of initial conditions and phase correlations
in the initial state (namely none). Since all quantities in (\ref{newsource}) are local functions of time, specified in terms of the mode functions (\ref{exactmode}) there is no need to resort to the nonlocal integral equation (\ref{nonl}), and the time evolving phase correlation ${\cal C}_{\bf k}$ need not be considered explicitly. Because of Eqns. (\ref{connection}) and (\ref{betabar}) the Schwinger pair creation amplitude is certainly contained 
in (\ref{newsource}). Since our objective is the derivation of an effective Markovian source term for the Boltzmann-Vlasov equation  
fields which are slowly varying in time we now introduce the second important ingredient in our approach, {\it i.e.} the uniform asymptotic expansion of (\ref{newsource}) for weak and slowly varying electric fields. 

In order to motivate the introduction of this asymptotic expansion
observe that for a constant electric field each time derivative 
of the mode function $f_{\bf k}$ brings with it a factor of $1/\lambda$. 
This can be made explicit by introducing a rescaled variable $v$ which 
is independent of the strength of the electric field, $u \equiv v \sqrt\lambda$
and rewriting the wave equation (\ref{waveq}) in the form,
\begin{equation}
\left({1\over \lambda^2}{d^2 \over dv^2} + v^2 + 1\right) f = 0\,.
\end{equation}
Next, when we allow the electric field to vary in time we can
consider the standard adiabatic expansion for the mode function 
in the time-varying field \cite{us1,us2}, 
\begin{eqnarray}
f_{\bf k} &\equiv &\sqrt{{\hbar\over 2 \Omega_{\bf k}}} 
\exp \left(-i\int^t \, dt'\,\Omega_{\bf k}(t')\right)\,,\nonumber\\
\Omega^2 &=& \omega^2 -{\ddot \Omega\over 2\Omega} +{3\over 4}
\left({\dot\Omega\over\Omega}\right)^2\nonumber\\
&=& \omega^2 -{\ddot \omega\over 2\omega} +{3\over 4}
\left({\dot\omega\over\omega}\right)^2 + \dots
\label{adbexp}
\end{eqnarray}
for which (\ref{adbmod}) is the lowest order term, corresponding to 
no derivatives in (\ref{adbexp}) and order $\lambda^0$ in the constant 
field case. The quasistationary or adiabatic approximation to the
mode equation is obtained by treating the derivative terms as
small compared to the leading order term, {\it i.e.}
\begin{equation}
{\ddot \omega\over \omega^3} \ll 1 \qquad {\rm and} \qquad  
{\dot\omega\over\omega^2} \ll 1\,.
\end{equation}
In the case of a general electric field this implies that the field is both
slowly varying and weak. For a constant electric field the adiabatic 
condition reduces to $\lambda \gg 1$.

Iterating the expansion to adiabatic order $q$, terms with 
$q$ time derivatives of the adiabatic frequency $\omega_{\bf k}$ in the 
general time varying field will appear together with terms with $q$ powers of $1/\lambda$ in the constant field case. Thus there is a one-to-one
correspondence between the asymptotic expansion of $|\beta_{\bf k}|^2$ and ${d \over dt}\vert{\beta}_{\bf k}\vert^2$ in powers of $1/\lambda$ in a constant electric field background to a local expansion of the current (\ref{curr2}) in higher time derivatives of the electric field in the general case. Now the transport approximation amounts to a truncation of this expansion at the lowest order required for a consistent backreaction dynamics. This is determined by the order of the backreaction equation, $j = -\dot E =\ddot A$, which is second order in time. Thus we should expand the particle number $|\beta|^2$ only to second order {\it i.e.} $1/ \lambda^2$, in 
order to match the asymptotic expansion of the current to the order of the backreaction equation for a weak, slowly varying electric field, 
self-consistently determined by solving the Maxwell-Vlasov system.
To retain higher orders than this in the current would not be correct 
mathematically, since these higher orders would also involve higher 
derivatives of $E$ in the general time varying electric field, and such 
terms can never be calculated correctly by the constant $E$ approximation
of (\ref{newsource}). At adiabatic order $2$ the only effect of approximating
the source term for a slowly varying electric field by (\ref{newsource}),
evaluated in a constant field is the absence of the $\dot E$ term in the current (\ref{curr}) generated by the adiabatic
expansion (\ref{adbexp}). This term is responsible for charge
renormalization in the mean field theory \cite{us1,us2}. Hence for comparison
between the mean field evolution and that of the Vlasov-Maxwell
system one must specify the scale of the renormalized charge of the mean field theory by some other criterion, or it will differ in general
from the classical charge appearing in the Vlasov equation
by a finite renormalization. This precise correspondence we fix by
a linear response analysis in Section VI.

Even if we could calculate higher order terms (by calculating
the source term in some other time varying background,
for example) it would not be correct physically to include them
since they would change the order of the Maxwell equation $\dot E = -j$ 
by making $j$ a function of higher derivatives of $E$ and thereby
introduce unphysical high frequency runaway solutions, not present in the underlying microscopic quantum field theory, in a manner similar to the
higher derivative Lorentz radiation reaction force. The important
physical point here is that the order of the backreaction equation
for time-varying electric fields determines the order of the
asymptotic expansion we should use for the current, in the
limit of weak, slowly varying electric fields, which is the only
limit in which such a replacement in the current is justified.
The fact that the leading order asymptotic expansion of the constant
field adiabatic particle number is already $1/\lambda^2$ (as we
shall see shortly) which is the highest order we need to go in the expansion,
justifies the use of the constant field expression (\ref{newsource}),
evaluated to this asymptotic order, for the local source term in the Markov limit of the quantum Vlasov equation. If the higher order terms in the expansion are numerically significant, then that is the signal that we must abandon the Boltzmann-Vlasov description entirely and return to the 
underlying field theory without the possibility of making any simple
transport approximation to the self-consistent backreaction problem.

The key point is that we require an asymptotic expansion of 
the mode functions and adiabatic particle number source term in 
(\ref{newsource}) in powers of $1/\lambda$
that is {\it uniformly} valid in time (and longitudinal momentum $k$),
in order that the exponentially small Schwinger amplitude $\bar\beta$
which is the only secular effect of particle creation which survives 
as $t\rightarrow \infty$ will not be lost in the expansion.
This condition is {\it not} satisfied by the naive asymptotic expansion
of $f$ in simple exponential functions such as (\ref{adbexp}).
What apparently has not been so generally
well recognized is that this failure of the usual adiabatic expansion
to capture exponentially small (but secular) particle creation effects 
is due to the nonuniformity of the naive asymptotic expansion with 
respect to the limits $t \rightarrow \pm \infty$. This limitation
can be removed by an asymptotic approximation uniformly valid
everywhere on the real time axis. 

In the case at hand, the asymptotic expansion of the solutions of Eqn. (\ref{waveq}) uniformly valid everywhere on the real time axis
have been given by Olver \cite{Olv}. Converting to the notations of the 
present paper, Olver's result may be written in the form,
\begin{equation}
f_{(+)\bf k}(t) \simeq e^{-{\pi\lambda\over 4}} 
\sqrt{2\pi\over \omega_{\bf k}} \left[1 +\sum_{s=1}\gamma_s {2^{s-1}\over(i\lambda)^s} \right]
\left\{ z^{1\over 4} {\rm Ai} (z) \sum_{s=0}{{\cal P}_{2s}
\over(i\lambda)^{2s}} + z^{-{1\over 4}} {\rm Ai'} (z)
\sum_{s=0}{{\cal Q}_{2s +1} \over(i\lambda)^{2s +1}}\right\}\,,
\label{asyairy}
\end{equation}
where the coefficients ${\cal P}_{2s}$ and ${\cal Q}_{2s + 1}$ are certain functions of $v={u\over\sqrt \lambda}$ given by
\begin{eqnarray}   
{\cal P}_0 (v) &=& 1\,,\nonumber\\
{\cal P}_2 (v)&=& - {(9v^4+ 249 v^2 -145) \over\ 1152  (v^2 + 1)^3} -
{7\ v\,(v^2+6) \over 1728 \xi (v^2+ 1)^{3\over 2}} + {455\over 10368 \,\xi^2}\,,\nonumber\\
{\cal Q}_1 (v)&=& -{i\over 24}\left[ {v(v^2+6) \over  (v^2 + 1)^{3\over 2}} - {5 \over 3\, \xi}\right]\,, \qquad etc.
\label{coairy}
\end{eqnarray}
the complex variables $\xi(v)$, $w$ and $z$ are defined by
\begin{eqnarray}
\xi(v) &\equiv & {\Theta_{\bf k}\over \lambda} + {i\pi\over 4}\nonumber\\
&=& {1\over 2} v {\sqrt{v^2 + 1}} + {1\over 2} \ln
\left( v + {\sqrt{v^2 + 1}}\right) + {i\pi\over 4}\qquad {\rm and}\nonumber\\
w&\equiv & -\lambda \xi \equiv {2i \over 3} z^{3\over 2}\,,
\label{defxiwz}
\end{eqnarray}
where $z$ is defined in the plane cut along the positive imaginary
axis by
\begin{equation}
z= e^{-i\pi} \left[ {3\over 2} \left(\Theta_{\bf k} + {i\pi\lambda\over 4}\right)\right]^{2\over 3}
\label{defz}
\end{equation}
and the $\gamma_s$ are the numerical constants,
\begin{equation} 
\gamma_1=-{1\over 24}\,,\qquad
\gamma_2={1\over 1152} \qquad {\rm etc.}
\end{equation}
With the definition of the phase of $z$ according to (\ref{defz}) the complex argument of the Airy functions in (\ref{asyairy})
varies along the contour depicted in Fig. 1, as the real time $t$ 
or $u$ ranges from $-\infty$ to $+\infty$.


\epsfxsize=11cm
\epsfysize=12cm
\vspace{1cm}
\centerline{\epsfbox{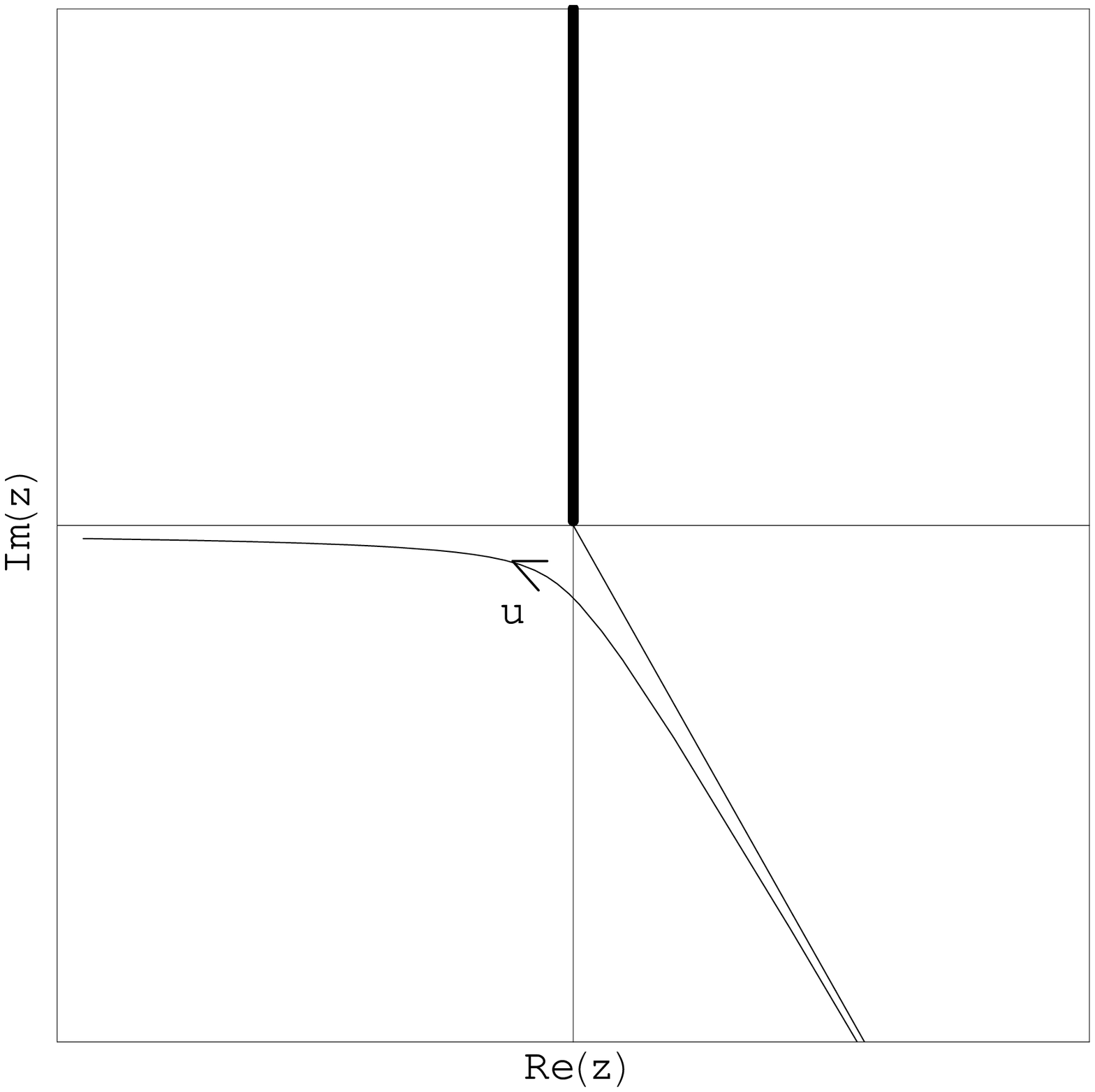}}
{FIG. 1. \small{The contour in the complex $z$ plane along
which the argument of the Airy functions in (\ref{asyairy}),
(\ref{asydairy}), (\ref{asybet}), and (\ref{lowadb}) are to
evaluated. The cut of the $2\over 3$ root appearing in (\ref{defz}) 
is taken along the positive imaginary $z$ axis from $0$ to $i\infty$. 
Following (\ref{defxiwz}) the corresponding contour in the complex
$w$ plane is a straight horizontal line displaced from the real $w$
axis into the lower half $w$ plane by $\pi\lambda\over 4$, with Re $w$ decreasing as $t$ increases.}} 
\vspace{1cm}
 
The terms we have written here explicitly determine the
uniform asymptotic expansion of $f_{(+)\bf k}$ up to order $1/\lambda^2$. 
Since we are interested only in the lowest nonvanishing order in the expansion
we could retain only the lowest order term in (\ref{asyairy}),
substitute it into (\ref{newsource}) to obtain the
lowest order source term in the Vlasov equation directly. 
Some care is required in this procedure since the argument of the Airy
functions depends on $\lambda$ through (\ref{defz}) and the equations of motion (\ref{abeom}) will not be satisfied unless both sides of the equation are expanded consistently to the same order in $1/ \lambda$. 
For this reason it is useful to retain one higher order in the asymptotic expansion than would seem necessary at first sight, in order to have a nontrivial check on the algebra via the equations of motion.

The corresponding asymptotic expansion for the time derivative
of the mode functions uniformly valid on the real axis is:
\begin{equation}
\dot f_{(+)\bf k}(t) \simeq i\,e^{-{\pi\lambda\over 4}} 
\sqrt{2\pi\omega_{\bf k}} \left[1 +{1\over 2}\sum_{s=1}\gamma_s {2^s\over(i\lambda)^s} \right]\left\{ z^{1\over 4} {\rm Ai} (z) 
\sum_{s=0}{{\cal P}_{2s + 1} \over(i\lambda)^{2s + 1}} 
+ z^{-{1\over 4}} {\rm Ai'} (z) \sum_{s=0}{{\cal Q}_{2s}
\over(i\lambda)^{2s}}\right\}\,,
\label{asydairy}
\end{equation}
where the coefficient functions are
\begin{eqnarray}
{\cal Q}_0(v)&=& 1\,,\nonumber\\
{\cal Q}_2(v) &=& {(15v^4+ 327 v^2 -143) \over 1152\ (v^2 + 1)^3}
+ {5\ v\,(v^2 - 6) \over 1728 \xi (v^2 + 1)^{3\over 2}} - {385\over 10368 \,\xi^2}
\,,\nonumber\\
{\cal P}_1(v) &=&  -{i\over 24}  \left[{v\,(v^2 - 6) \over  (v^2 + 1)^{3\over 2}}
+ {7 \over 3\, \xi}\right]\qquad etc.
\label{codairy}
\end{eqnarray}
From these expressions the uniform asymptotic expansion for the time dependent
Bogoliubov coefficient $\beta_{\bf k}(t)$ is easily
computed from its definition in (\ref{alpbet}), namely,
\begin{equation} 
\beta_{\bf k} \simeq \sqrt{\pi} e^{-{\pi\lambda\over 4}} e^{-i\Theta_{\bf k}}
\left[1 +{1\over 2}\sum_{s=1}\gamma_s {2^s\over(i\lambda)^s} \right]\left\{z^{1\over 4} {\rm Ai} (z) \sum_{s=0}{{\cal P}_s \over(i\lambda)^s} + z^{-{1\over 4}} {\rm Ai'} (z)\sum_{s=0}{{\cal Q}_s \over(i\lambda)^s}\right\}\,.
\label{asybet}
\end{equation}

Since we have shown by Eqn. (\ref{enerom}) that the particle number $|\beta_{\bf k}|^2$ is an adiabatic invariant to leading order in the time derivatives of the background, the lowest order $\lambda^0$ term in the asymptotic expansion must be absent from the particular linear combination in (\ref{asybet}). Indeed with $s=0$, ${\cal P}_0 = {\cal Q}_0 = 1$
and the symmetric linear combination of
Airy functions, $z^{1\over 4} {\rm Ai} (z)  + z^{-{1\over 4}} {\rm Ai'} (z)$,
is of order $\lambda^{-1}$, as is verified explicitly in
relations (\ref{AirHan}) and (\ref{hankasy}) below, by using (\ref{defxiwz}) and the further asymptotic expansion of these functions for $|z| \sim \lambda^{2\over 3} \rightarrow \infty$. Any other linear
combination of the same functions, and in particular the
antisymmetric combination, $z^{1\over 4} {\rm Ai} (z)  - z^{-{1\over 4}} {\rm Ai'} (z)$ is of order $\lambda^0$. Anticipating this result and substituting (\ref{coairy}) and (\ref{codairy}) into (\ref{asybet}), we obtain simply,
\begin{eqnarray}
\beta_{\bf k} &\simeq &\sqrt{\pi} e^{-{\pi\lambda\over 4}} e^{-i\Theta_{\bf k}}
\left\{\left[z^{1\over 4} {\rm Ai} (z)  + z^{-{1\over 4}} {\rm Ai'} (z)\right]
+ {i\over 4} \left[z^{1\over 4} {\rm Ai} (z)  - z^{-{1\over 4}} {\rm Ai'} (z)\right] \left[{u\over (u^2+\lambda)^{3\over 2}} + {1\over 3w}\right] 
\right\}\nonumber\\
& & \qquad + {\cal O} (\lambda^{-2})\,,
\label{lowadb}
\end{eqnarray}
correct to the leading nonvanishing order, $\lambda^{-1}$. 
Squaring (\ref{lowadb}) and taking its time derivative gives the asymptotic
approximation to the effective source term defined in (\ref{newsource}). 
Since $w$ and $z$ are 
functions of $u$ and $\lambda$ (equivalently, $v$ and $\lambda$) which depend 
only on the {\it kinetic} momenta $p(t)$ and $p_{\perp}$ through
(\ref{rescaled}), the effective source term (for vacuum intitial conditions
at $t= -\infty$) may be written in the form,
\begin{equation}
S_{vac}(p, p_{\perp};E) = eE {\partial\over \partial p} \vert\beta (p, p_{\perp})\vert^2\,,
\label{finsource}
\end{equation}
where 
\begin{equation}
\vert\beta (p, p_{\perp})\vert^2 = \pi e^{-{\pi\lambda\over 2}}\left\vert
\left[z^{1\over 4} {\rm Ai} (z)  + z^{-{1\over 4}} {\rm Ai'} (z)\right]
+ {i\over 4} \left[z^{1\over 4} {\rm Ai} (z)  - z^{-{1\over 4}} {\rm Ai'} (z)\right] \left[{u\over (u^2+\lambda)^{3\over 2}} + {1\over 3w}\right] 
\right\vert^2
\label{betkin}
\end{equation}
is that function of $p$ and $p_{\perp}$ determined by the
subsitutions (\ref{rescaled}) together with the definitions (\ref{defxiwz})
and (\ref{defz}). Using the relations,
\begin{eqnarray}
eE {\partial\over \partial p} \left[z^{1\over 4} {\rm Ai} (z)  \pm
z^{-{1\over 4}} {\rm Ai'} (z)\right] &=& \pm i \omega 
\left[z^{1\over 4} {\rm Ai} (z)  \pm z^{-{1\over 4}} {\rm Ai'} (z)\right]
\nonumber\\
& & \qquad +  {i \omega \over 4 z^{3 \over 2}}\left[z^{1\over 4} {\rm Ai} (z)  \mp z^{-{1\over 4}} {\rm Ai'} (z)\right]
\end{eqnarray}
which follow from the definition (\ref{defz}) and ${\rm Ai''}(z) = z {\rm Ai}(z)$, the differentiation in (\ref{finsource}) can be carried
out explicitly with the result,
\begin{eqnarray}
&&\quad S_{vac}(p, p_{\perp};E) = \pi \vert eE\vert e^{-{\pi\lambda\over 2}}{u\over
u^2 + \lambda} \Bigg\{\vert z^{1\over 4} {\rm Ai} (z)\vert^2 -
\vert z^{-{1\over 4}} {\rm Ai'} (z)\vert^2 \nonumber\\
&&\qquad + {1\over 6} {\rm Im}\left[\left(z^{1\over 4} {\rm Ai} (z)\right)^*z^{-{1\over 4}} {\rm Ai'} (z)\right]
\left({\Theta_{\bf k}\over 3\vert w\vert^2} + {u^3 \over\lambda (u^2 + \lambda)^{3\over 2}}\right) -{1\over 48}\vert z^{1\over 4} {\rm Ai} (z) - z^{-{1\over 4}} {\rm Ai'} (z) \vert^2 \nonumber\\
&&\times\left[{\pi\lambda\over \vert w\vert^2} + {6(2u^2 -\lambda)\over (u^2 + \lambda)^3} + {u^3 \Theta_{\bf k}\over 3 \lambda \vert w\vert^2 (u^2 + \lambda)^{3\over 2}} + {35\over 18 \vert w\vert^4}
\left( {\pi^2\lambda^2\over 16}- \Theta_{\bf k}^2 \right) + {1\over 18 \vert w\vert^2}\right]\Bigg\}\,,
\label{thesource}
\end{eqnarray}    
where we have neglected all terms of order $\lambda^{-3}$
and higher within the curly brackets.

The expressions (\ref{finsource}), (\ref{betkin}) and (\ref{thesource})
are the main results of this paper. In order to understand the physics of particle creation that is captured in these expressions we compare the lowest order asymptotic expression for the adiabatic particle number (\ref{betkin})
with the analogous exact expression in terms of parabolic cylinder functions for constant external electric field. The results are plotted in Figs. 2 through 4 for $\lambda = 1, 2$ and $10$ respectively. We see that the asymptotic expansion in terms of Airy functions reproduces the behavior 
of the adiabatic particle number quite accurately, even for moderately small $\lambda$ of order one. The other important feature to notice about these figures is the relatively sharp increase in particle number right around 
$u = p = 0$. The transients after this particle creation event then settle down to the value $|\bar\beta|^2 = \exp(-\pi\lambda)$ which is
independent of the initial longitudinal momentum.

Thus, the exponentially small Schwinger particle creation effect
is captured very well by the {\it leading} order term in the uniform asymptotic expansion of $|\beta|^2$. Notice that the uniform asymptotic
expansion for the source term works very well even at the expected
limit of its validity at $\lambda =1$. 
As a mathematical aside we remark that the exponentially small 
contribution to an adiabatic invariant quantity such as the particle number
${\cal N}_{\bf k}$ has been studied by various authors and bounds
obtained in the general case \cite{Gol}. However, for this particular case
of constant electric field and mode equation (\ref{waveq}) leading
to Weber parabolic cylinder functions, it has apparently not been
noticed that the asymptotic expansion of the solutions of this
equation, uniformly valid on the real time axis, allows one to
calculate the exponentially small secular change in the adiabatic
invariant ${\cal N}_{\bf k}$ analytically. 
The same observation could clearly be generalized to
other differential equations for which uniform asymptotic
expansions are known.


\epsfxsize=16cm
\epsfysize=18cm
\vspace{4cm}
\centerline{\epsfbox{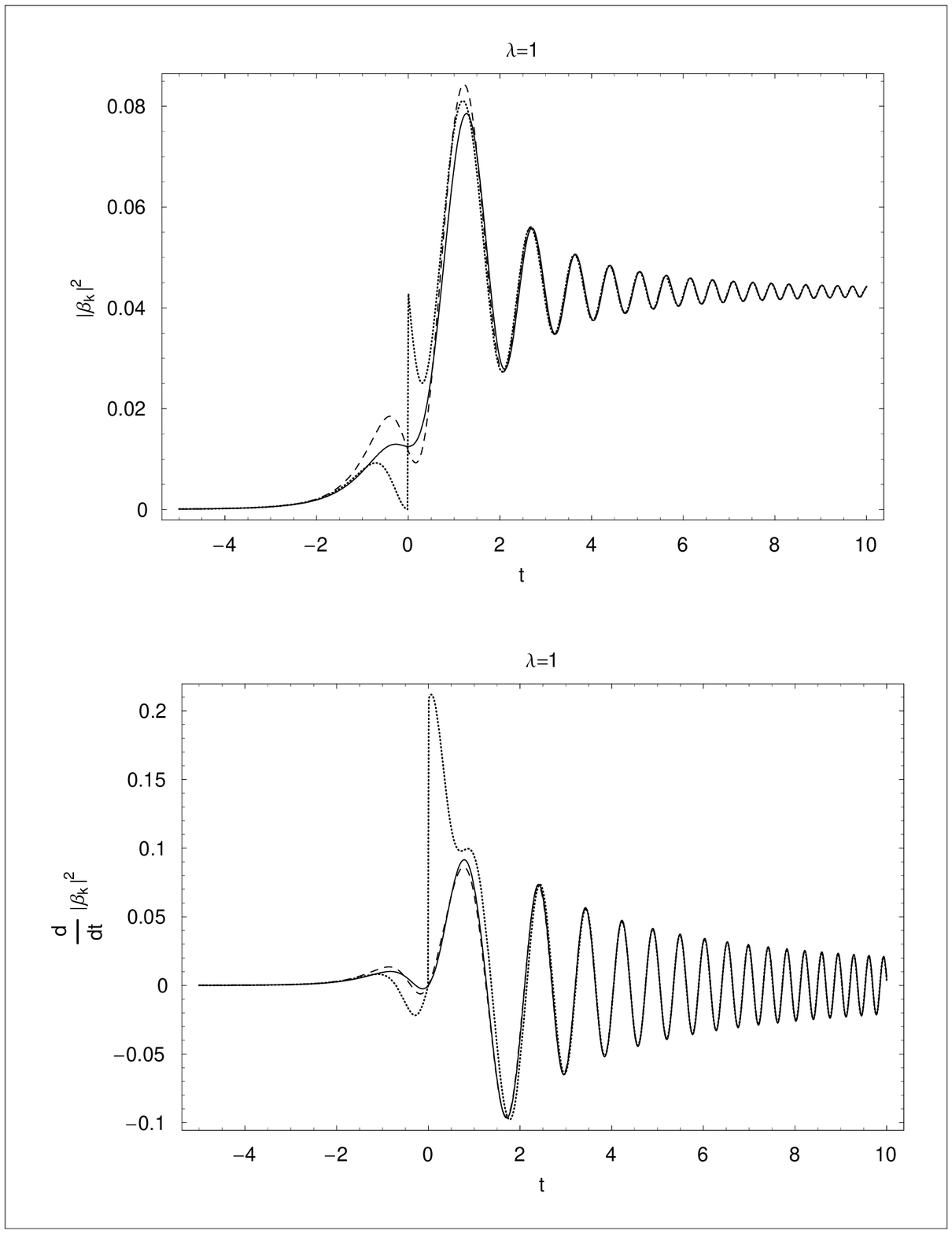}}
\vspace{1cm}
{FIG. 2. \small{The exact (solid curve), uniform (dashed curve) and 
adiabatic step function (dotted curve) asymptotic expansions of the
adiabatic particle number and its time derivative for a
constant electric field with $\lambda =1$ and $k=0$. The particle numbers approach the same value $e^{-\pi}= 0.0432$ as $t\rightarrow\infty$, although each ${\cal N}_k$ experiences a sharp rise at a different time,
{\it viz.} near zero kinetic momentum, $p=k+eEt \simeq 0$. The delta
function at $t=0$ in the dotted curve of the second figure obtained
from differentiating (\ref{betappr}) is not shown.}} 

 
\epsfxsize=16cm
\epsfysize=18cm
\vspace{4cm}
\centerline{\epsfbox{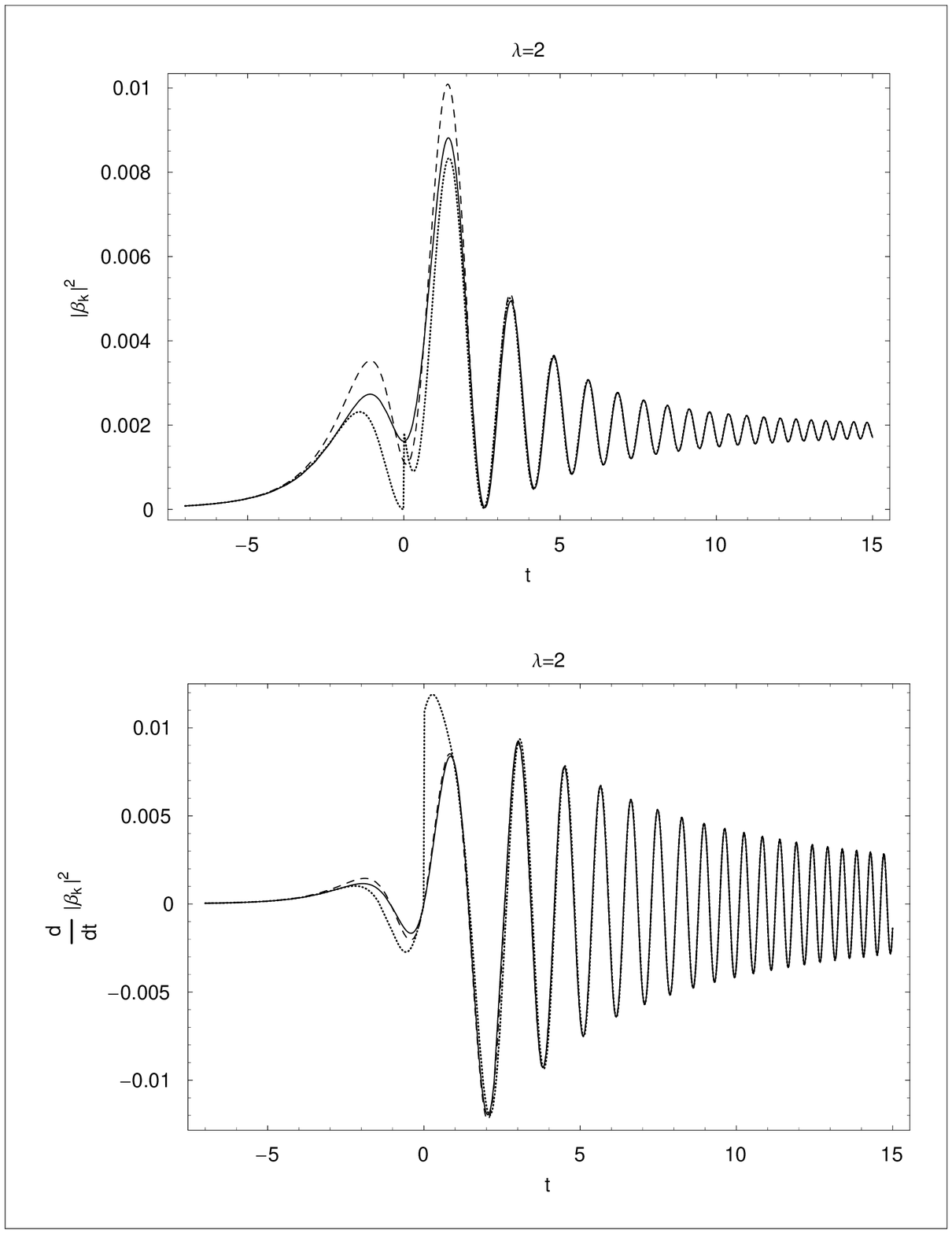}}
\vspace{1cm}
{FIG. 3. \small{Same as Fig. 2 but for $\lambda =2$. The particle numbers approach the same value $e^{-2\pi}= 0.00187$ as $t\rightarrow\infty$, although each ${\cal N}_k$ experiences a sharp rise at a different time,
{\it viz.} near zero kinetic momentum, $p=k+eEt \simeq 0$.}} 


\epsfxsize=16cm
\epsfysize=18cm
\vspace{4cm}
\centerline{\epsfbox{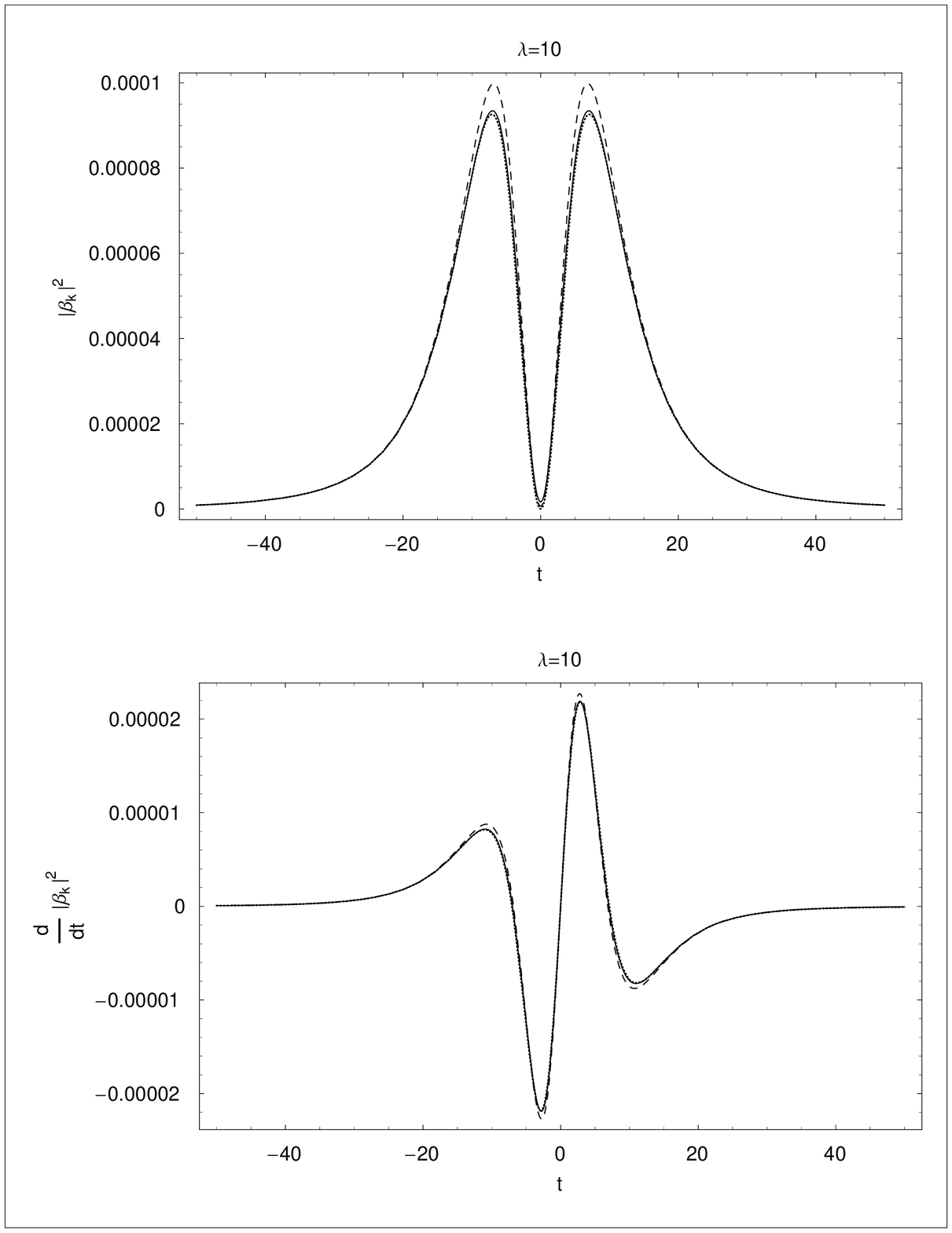}}
\vspace{1cm}
{FIG. 4. \small{Same as Figs. 2 and 3 but for $\lambda =10$. In this case the magnitude of the step at late times, $e^{-10\pi}= 2.27 \times 10^{-14}$ is much smaller than the transient effects visible in the plot, and all three curves are very nearly (anti)symmetric around $t=0$, showing
that a nearly equal number of particles is created and destroyed. As in the previous figures the delta function at $t=0$ in the dotted curve of the second figure is not shown.}} 
\newpage

The sharpness of the creation event at $u=0$ is clearly determined by
the wave equation (\ref{waveq}) to be $\Delta u \sim \lambda^{1\over 2}$ or
\begin{equation}
\Delta t \sim {\sqrt{p_{\perp}^2 + m^2c^2}\over eE} \equiv \tau_{cl}\,
\end{equation}
which is the time scale for the growth of a sizable fraction of the
final antiparticle amplitude
in the quantum wave function. This time scale (which is also
the time scale for the classical acceleration by the
electric field to bring a charged particle to relativistic velocities)
must be long compared to the quantum phase coherence time 
$\tau_{qu}$, in order for the creation process to be described 
by a local approximation to the nonlocal Vlasov equation (\ref{nonl}), 
{\it i.e.}
\begin{equation}
\tau_{cl} \sim \lambda \tau_{qu} \gg \tau_{qu}\, . 
\end{equation}

Hence the Markov limit of the Vlasov equation requires weak
electric fields $\lambda \gg 1$ which is what we have assumed in the
uniform adiabatic expansion of the source term.
Conversely, if we consider the opposite limit where the electric field
is strong, so many particles are created so rapidly in time that 
the individual particle creation events cannot be distinguished 
one from another during the quantum coherence time $\tau_{qu}$.
It is clear that in this case significant wave amplitude coherence
during the creation process can be expected and we cannot hope to 
approximate the effects of such copious and coherent particle creation by a Boltzmann-Vlasov source term local in time, which takes no account of the 
prior time history. Indeed in this strong field limit these ``particles" 
are not particles at all in the usual sense but are more accurately to be thought of as coherent wave amplitudes which lie outside of any
classical or semiclassical kinetic particle description.

Restricting ourselves then to weak fields these coherence effects
do not need to be considered explicitly and are built into the
initial conditions of the vacuum at $t= -\infty$ once and for all.
However in analyzing the particle creation process in a 
constant field and deriving (\ref{thesource}) 
we have also assumed that the electric field does not vary over
the typical time of the variation of ${\cal N}_{\bf k}$. Thus
in order to use (\ref{thesource}) in situations
involving a time evolving electric field we also require that
its time scale of variation $\tau_{pl}$ be much larger than the time
scale of the creation event, {\it i.e.},
\begin{equation}
\tau_{pl} \gg \tau_{cl}\,.
\label{secineq}
\end{equation}
If this second inequality holds then it should be possible to
coarsen our time resolution still further by not attempting
to resolve the time scale $\tau_{cl}$. On these
still longer time scales it becomes reasonable to
approximate the sharp growth of the antiparticle amplitude near
$u = p_z = 0$ as a step function, provided only that
we account for the integrated value of the step from $-\infty$
to $+\infty$. This is what we wish to explain next.

Let us first reiterate that the uniform asymptotic expansion
in terms of Airy functions is indeed essential to capturing
the step explicitly in Figs. 2-4 and that the Schwinger effect
is lost completely if a naive WKB expansion in powers of $1/\lambda$
is used instead. This may be seen explicitly by taking
the large $\lambda$ asymptotics of the Airy functions in (\ref{lowadb}).
To this end we note the Airy functions may be represented in terms of Hankel functions of the first kind,
\begin{eqnarray}
z^{1\over 4} {\rm Ai} (z) &=& {1\over 2 \sqrt{2}}e^{5\pi i \over 12}\, 
w^{1\over 2} H^{(1)}_{1\over 3}(w)\nonumber\\
z^{-{1\over 4}} {\rm Ai'} (z) &=& {1\over
2 \sqrt{2}}e^{-{5\pi i \over 12}}\, w^{1\over 2} H^{(1)}_{2\over 3}(w)
\label{AirHan}
\end{eqnarray}
with the branch cut of $w^{1\over 2}$ along the negative $w$ axis,
and $w$ ranging from $+\infty -{i\pi\lambda\over 4}$ to
$-\infty -{i\pi\lambda\over 4}$ along the horizontal contour
displaced by $ -{i\pi\lambda\over 4}$ from the real axis, as
$u$ ranges from $-\infty$ to $+\infty$, according to (\ref{defxiwz}).
Taking the large $\lambda$ limit is equivalent to
taking the large $|w|$ limit of the Hankel functions, which
depends critically on the phase of $w$. This phase depends in
turn on the sign of $u$ from (\ref{defxiwz}). 
When $u <0$, then $|{\rm arg}\, w| < \pi$
and we can use the standard asymptotic expansion of the Hankel functions,
\begin{equation}
w^{1\over 2} H_{\nu}^{(1)} (w) \simeq {2\over \sqrt\pi} \exp \left(i 
w - i{\pi\over 2}\nu - i{\pi \over 4}\right)\left\{ 1 -{1\over 2iw}
{\Gamma(\nu + {3\over 2})\over \Gamma(\nu - {1\over 2})} +\dots\right\}\,,
\qquad -\pi < {\rm arg}\, w < \pi\,,
\label{hankasy}
\end{equation}
for large $|w|$ to find 
\begin{equation}
\beta_{\bf k} \sim {e^{iw}\over w} \rightarrow 0 \qquad {\rm as} \qquad
t\rightarrow -\infty\,.
\end{equation}
Since $w = -\lambda \xi$ is linear in $\lambda$ this shows that
the linear combination of Airy or Hankel functions in $\beta_{\bf k}$ is 
of order $\lambda^{-1}$ and contains no $\lambda^0$
term, as stated above.

Thus if $t \rightarrow -\infty$ with $eE$ fixed or if $\lambda \rightarrow
\infty$ with $k + eEt <0$ fixed, the adiabatic particle number vanishes. 
On the other hand if $u \rightarrow +\infty$ we cannot use (\ref{hankasy})
directly because arg $w \rightarrow -\pi$ in this limit, and
the condition on the phase is not satisfied.
Instead we must first use the connection formula,
\begin{equation}
H_{\nu}^{(1)} (w)  = {\sin 2\pi \nu \over \sin \pi \nu} H_{\nu}^{(1)} (e^{i\pi}w) + e^{-i\pi\nu} H_{\nu}^{(2)} (e^{i\pi}w)
\end{equation}
to bring the phase of $w' = e^{i\pi} w$ into the proper range in order
to apply (\ref{hankasy}). Then we find that the $e^{iw}$ terms from
$H_{\nu}^{(1)}$ again vanish like $e^{iw}|w|^{-1}$ for large $|w|$, but that now there remains in addition the opposite frequency $e^{-iw}$ term which gives
\begin{equation}
\beta_{\bf k} (t) \rightarrow -i e^{-{\pi\lambda\over 4}} e^{-iw} e^{-i\Theta_{\bf k}} = -i e^{-{\pi\lambda\over 2}} = \bar\beta
\end{equation}
which is finite as $t \rightarrow +\infty$ with $eE$ fixed.
As $\lambda \rightarrow \infty$ with $k + eEt >0$ fixed
this term is exponentially small compared to the ordinary $\lambda^{-1}$
contribution. 

In this way the uniform asymptotic expansion in terms of Airy or Hankel functions which contains the exponentially small Schwinger 
particle creation becomes non-uniform in time, depending on the sign of 
$k + eEt$, if the further asymptotic expansion of these functions in
terms of exponentials $\exp(\pm iw)$ is taken. Only the
uniform expansion in (\ref{asyairy}) and (\ref{asydairy})
can capture the particle creation event, and Figs. 2-4 show
that this it does quite accurately even at the lowest nonvanishing
order of the expansion. This exercise
in asymptotic expansions as well as the explicit behavior in time of the
adiabatic particle number in Figs. 2-4 does shows that we might
try the simple adiabatic expansion of $\beta_{\bf k}$ according to
(\ref{adbexp}), but that we must then add back {\it by hand} the exponentially small step $\bar\beta$ in the vicinity of the creation event at $k + eEt \simeq 0$, {\it i.e.}
\begin{equation}
\beta_{\bf k} \approx \beta_{\bf k}^{adb} + \theta(u) \bar\beta\,,
\end{equation}
where
\begin{equation}
\beta_{\bf k}^{adb} \simeq {i e^{-2i\Theta_{\bf k}} u \over 4 (u^2 + \lambda)^{3\over 2}} + {\cal O}\left({1\over \lambda^2}\right)
\end{equation}
is the result obtained by substituting the lowest order of the standard adiabatic approximation for the mode functions (\ref{adbexp}), 
rather than Olver's uniform expansion in terms of Airy
functions. The Heaviside step function could be replaced by any
smooth function with the correct limits at $t \rightarrow \pm \infty$.
The point is that if the second inequality (\ref{secineq}) holds,
then in multiparticle collective quantities such as the
mean current, integrations over large ranges of kinetic momenta are
involved, and it makes little difference whether the continuous
rise in each individual mode's particle number on the momentum scale
$\tau_{cl}/eE$ is taken into account, provided only that the integral 
over all momenta accurately describes which modes have gone
through the creation process. It is only this fact and the
second inequality involving the collective time scale of the plasma that can justify replacing the continuous rise of ${\cal N}_{\bf k}$ by a step function. 

In this admitedly rather crude approximation the function $\beta (p, p_{\perp})$ of (\ref{betkin}) in terms of Airy or Hankel functions
is replaced by
\begin{equation}
|\beta (p, p_{\perp})|^2 \approx {(eE)^2 p^2 \over 16\omega^6}
- {eE p\over 2 \omega^3}\exp\left(-{\pi \lambda\over 2}\right)
\theta(u) \cos (2\Theta)
+ \exp\left(-\pi\lambda\right)\theta (u)
\label{betappr}  
\end{equation}
in terms of elementary functions. This approximation to (\ref{betkin}) is compared to the uniform asymptotic expansion in the dotted
curves of Figs. 2-4, where it is
observed that it works better than might have been expected, except
for the region near the creation event $u \approx 0$ where it is
clearly inaccurate. The delta function obtained by differentiating
the last two terms of (\ref{betappr}) is not shown in the second
halves of Figs. 2-4. Notice that the oscillations in these
figures are well represented by the $\cos (2\Theta)$ term in (\ref{betappr}), which may be interpreted as the interference between the usual adiabatic 
phase oscillations and the relatively sudden particle creation event.
Thus we see that for numerical purposes it is probably sufficient to
use the approximate form of $|\beta (p, p_{\perp})|^2$ in (\ref{betappr})
for all $p$, except those in a band of size several units of $\sqrt{p_{\perp}^2 + m^2}$ centered at the origin where the sharp (but continuous) rise of particle number takes place. When one is integrating over
a region of $p$ or $t$ that is large compared to the time scale
$\tau_{cl}$ over which the rise in particle number takes place,
the crude approximation of this rise by a step function and its
derivative by a delta function may be sufficient, provided only
that their coefficient is fixed by the Schwinger formula, as in 
(\ref{betappr}). On the other hand, in the region of $p \simeq 0$
the true behavior is certainly not discontinuous on the scale $\tau_{cl}$ and
the more accurate form (\ref{betkin}) in terms of Airy or Hankel functions should be used for moderately strong electric fields. 

We have now succeeded in our main purpose, namely to analyze the time 
structure of the quantum particle creation process in the adiabatic number 
basis, and to capture that particle creation event by means of a uniform
asymptotic expansion of the exact wave functions of the constant
electric field background, without any need to analytically continue
or approximate the nonlocal integral in (\ref{nonl}). Because of the reasoning
earlier in this section we can proceed to identify the time rate of
change of the adiabatic particle number in the lowest order of this
uniform asymptotic expansion given by (\ref{thesource}) or the time derivative of (\ref{betappr}) as the effective source term in the Vlasov equation which describes quantum particle creation in slowly varying electric fields, starting from vacuum initial conditions
at $t = -\infty$.

One point that still requires some discussion is the effect
of changing the initial conditions from vacuum at $t= -\infty$
to those at some finite time $t_0$. Indeed, the comparison of the 
effective source term in the Vlasov
description with the mean field evolution in the next section requires
that the initial conditions be specified at a finite initial time $t_0$, 
not at $-\infty$. This means that the effective source term which is given by
Eqns. (\ref{thesource}) will differ from the
actual source term in its dependence on the initial data and the
correlations (or lack of them) in the initial state. To the
extent that a Markovian approximation to the source term is justified
and dephasing is efficient we expect that the memory effects of the 
initial conditions will be washed out on the time scale of significant 
particle creation, and therefore that the initial conditions will affect 
only the transient behavior of the evolution for times close to $t_0$.
This can be checked in more detail.

To examine the transient effects of the initial conditions let us consider arbitrary initial data on the mode functions 
$f_{\bf k}(t_0)$ and $\dot f_{\bf k}(t_0)$, subject only to the Wronskian
condition (\ref{Wron}) and finite initial energy density. 
The general solution of $f_{\bf k}(t)$ in a constant electric field is 
a linear combination of $f_{(\pm)\bf k}$,
\begin{equation}
f_{\bf k}(t)=A_{\bf k}(t_0) f_{(+)\bf k}(t) 
           + B_{\bf k}(t_0) f^{\ast}_{(+)\bf k}(t)\, .
\label{exactsol}
\end{equation} 
By using the Wronskian condition on the mode functions we can solve
for the coefficients, $A_{\bf k}(t_0)$ and $B_{\bf k}(t_0)$ in terms of the initial conditions on the mode functions in the form,
\begin{eqnarray}
A_{\bf k}(t_0)&=&  i(\dot{f}_{\bf k}(t_0) f^{\ast}_{(+)\bf k}(t_0) 
- f_{\bf k}(t_0) \dot{f}^{\ast}_{(+)\bf k}(t_0) ) \nonumber \\
B_{\bf k}(t_0)&=&  i(f_{\bf k}(t_0) \dot{f}_{(+)\bf k}(t_0) 
- \dot{f}_{\bf k}(t_0) f_{(+)\bf k}(t_0) ) \, .
\label{ABdef}
\end{eqnarray}
A specific example of initial data with finite energy density
is provided by the adiabatic vacuum initial conditions at $t=t_0$, {\it i.e.},
\begin{eqnarray}
f_{\bf k}(t_0)&=& {\tilde f}_{\bf k}(t_0) = \sqrt{{\hbar\over2 \omega_{\bf k}(t_0)}} \nonumber \\
\dot{f}_{\bf k}(t_0)&=& \dot{\tilde f}_{\bf k}(t_0) = \left[-i\omega_{\bf k}(t_0) +{\dot{\omega}_{\bf k}(t_0)
\over 2 \omega_{\bf k}(t_0)} \right]
{f}_{\bf k}(t_0) \, .
\label{initcond}
\end{eqnarray}
The second term in the time derivative of the mode function
is essential to insure finite initial energy density and
is nonzero for finite electric field at initial time $t_0$. It means
that a definite nonzero value of the pair correlation 
${\cal C}_{\bf k}$ is being assumed in the initial adiabatic vacuum state.

Our previous choice of vacuum initial conditions is recovered
if we let $t_0 \rightarrow -\infty$ so that $A_{\bf k} \rightarrow 1$,
$B_{\bf k} \rightarrow 0$ and $f_{\bf k}(t) \rightarrow f_{(+)\bf k}(t)$.
Retaining $t_0$ finite means that the general expression for
$\beta_{\bf k}(t)$ in (\ref{alpbet}) with the mode functions given
by (\ref{exactsol}) should be used so that 
\begin{eqnarray}
\beta_{\bf k}(t, t_0) = -iA_{\bf k}(t_0){\tilde f}_{\bf k}(t)
(\dot f_{(+)\bf k} + i \omega_{\bf k} f_{(+)\bf k})
-iB_{\bf k}(t_0){\tilde f}_{\bf k}(t)
(\dot f^{\ast}_{(+)\bf k} + i \omega_{\bf k} f^{\ast}_{(+)\bf k})\, .
\label{betAB}
\end{eqnarray}
Since $A_{\bf k}$ and $B_{\bf k}$ are also given in terms of $f_{(+)\bf k}$
by Eqns. (\ref{ABdef}) one can develop the uniform asymptotic
expansion for this $\beta_{\bf k}(t, t_0)$ using (\ref{asyairy}) and
(\ref{asydairy}), repeating the steps leading to Eqn. (\ref{betkin})
keeping $t_0$ finite. 

The resulting rather complicated expression for the source term will
depend on the electric field value at the initial time $t_0$.
This expression would incorporate the initial data of the actual mean field
evolution problem starting at $t_0$ more accurately than the simple 
choice of initial conditions, $A_{\bf k} =1$ and $B_{\bf k} = 0$
which we have used in the source term (\ref{thesource}).
A good probe of the effect of these transient terms is the electric current
which is plotted in Fig. 5 for $\lambda = 1$.
The early oscillations observed in the exact current are the effect of the initial conditions (\ref{initcond}). However the linear growth
with $t$ at late times can be understood from
the simple approximation to the particle creation
(\ref{betappr}) by a step function.


\epsfxsize=16cm
\epsfysize=12cm
\centerline{\epsfbox{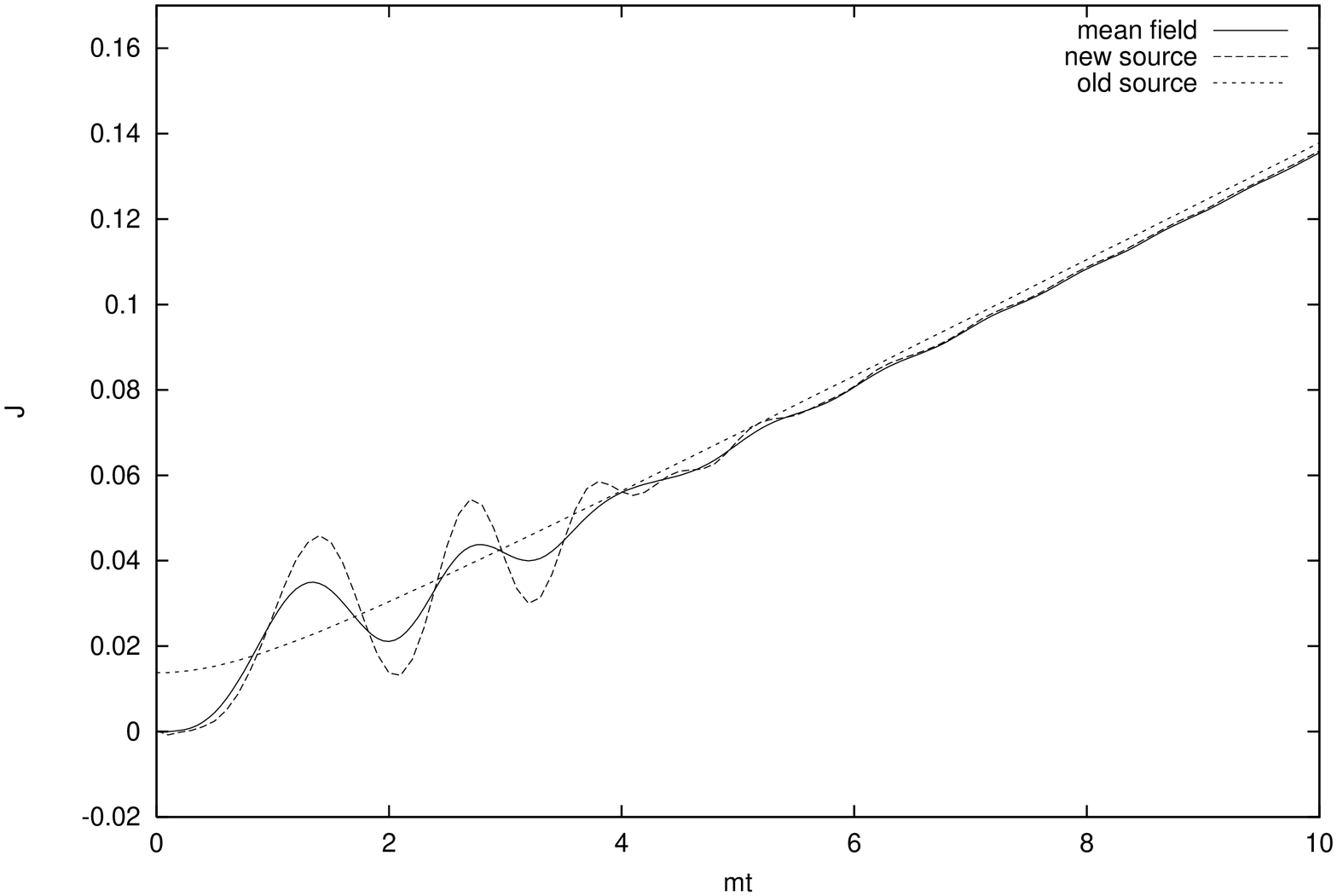}}
{FIG. 5. \small{The linear growth of the electric current with time in
the case of fixed constant background electric field $E=1$ and $e=1$. 
The three curves shown are the current of the exact mode functions, the uniform Airy approximation to them with initial conditions at $t_0 =0$ according to (\ref{exactsol}-\ref{initcond}), and the simple step function ansatz of Eqn.
(\ref{betappr}).}}
\vspace{1cm}

For if we start at $t=t_0$ 
(rather than at $t=-\infty$) with no initial particles present, 
then the actual current integrated over all longitudinal momenta at time $t$ is dominated by the conduction current $j_{cond}$
in (\ref{curr3}) and becomes
\begin{equation}
2e\exp\left(-\pi\lambda\right)\int\, {dk\over 2\pi} {k + eEt \over \omega_k}\, \theta (k + eEt) \theta (-k - eEt_0) \rightarrow 
{e^2E \over \pi}\exp\left(-\pi\lambda\right)\, (t-t_0)
\label{lincur}
\end{equation}
in one spatial dimension at late times, which grows linearly with the elapsed time $T = t-t_0$ since the initial vacuum state was prepared. This
is precisely the slope which is observed in all three curves in
Fig. 5 at late times. The second step function
involving $t_0$ is necessary because only modes with initially negative kinetic momentum can go through a creation event at $p(t) \simeq 0$ since
$p(t) = k + eEt$ is a monotonically increasing function of $t$ for
constant positive $E$. It is in fact present in the $A_{\bf k}$ and 
$B_{\bf k}$ of (\ref{ABdef}) through $f_{\bf k}(t_0)$ and 
$\dot f_{\bf k}(t_0)$ which involve the same parabolic cylinder mode functions and the similar behavior near $k + eEt_0 \simeq 0$ as observed in
Figs. 2-4. It is this dependence on the initial data which
provides just the momentum window in (\ref{lincur}) which we need to justify the replacement of the longitudinal
momentum integration $dk$ by the total elapsed time $eET$ in (\ref{rate}), and which led to Schwinger's result for the decay rate. Such an
understanding of the linear time divergence is possible only with
a detailed description of the time structure of the particle creation 
process as given here. 
 
The fact that the current in a constant electric field grows linearly
with time is important for another reason. For it shows that backreaction 
must eventually be taken into account, and that simple perturbation theory
must break down at late enough times for any nonzero $eE$, no matter how small. These backreaction effects can be taken into account only by a
systematic resummation of perturbation theory, such as the large $N$ expansion
advocated in refs. \cite{us1}-\cite{us3}, or by the solution of the
Vlasov-Maxwell system of equations, valid when the inequalities
of time scales $\tau_{qu} \ll \tau_{cl} \ll \tau_{pl}$ hold. 

When the electric fields are very weak fields ($eE \ll { m^2 c^3/\hbar}$), particle creation is negligible, the linear slope in Fig. 5 is very
small and even in backreaction the electric field will hardly change
at all with time. In this case essentially all the effects on
moderate time scales will be transient effects and one should
reatin the initial condition information in $A_{\bf k}$ and $B_{\bf k}$.
In moderately strong electric fields 
($eE \simeq { m^2 c^3/\hbar}$) where particle creation is significant
Fig. 5 shows that the transient effects of the initial data become unimportant
before long and one might just as well
use the simpler expression for the source term with $A_{\bf k} =1$ 
and $B_{\bf k} = 0$, derived previously. This is equivalent to 
replacing the electric field value
the particles feel at the actual time of creation by one assumed
to have been constant for times long before the creation takes place.
In that case the source term does not depend
on the value of the electric field at the initial time $t_0$,
which again is reasonable provided $\tau_{pl} \gg \tau_{cl}$. 
It is the quasistationary, Markov approximation for the 
source term in (\ref{thesource}) or (\ref{betappr})
$A_{\bf k} =1$ and $B_{\bf k} = 0$ that we compare to the actual backreaction evolution of mean field theory in the next section.

We conclude this section by remarking on the relationship between the local source term (\ref{newsource}) or its asymptotic expansion,
(\ref{thesource}) and the general nonlocal form (\ref{nonl}) derived in Section II. For a constant electric field starting from vacuum initial conditions (\ref{newsource}) and (\ref{nonl}) must be identical of course.
If, following Rau \cite{Rau} one neglects the Bose enhancement
factor $1 + 2 {\cal N}_{\bf k}$ and changes variables from $t'$ to
$\lambda x \equiv 2 \Theta_{\bf k}(t') - 2\Theta_{\bf k}(t)$ then
the integral in (\ref{nonl}) may be rewritten in the form,
\begin{equation}
{d \over dt}{\cal N}_{\bf k} = {eE p\over 4 \omega^2} \int_{-\infty}^0\, dx\,
{{\rm sinh} \varphi (x) \over {\rm cosh}^3 \varphi (x)} \cos (\lambda x)
\label{Rausource}
\end{equation}
where 
\begin{equation}
u' = \epsilon { k + eE t' \over \sqrt{\vert eE\vert}} = \sqrt{\lambda}\ {\rm sinh}
\varphi (x)
\end{equation}
is given implicitly as a function of $x$ by the relation,
\begin{equation}
{\rm sinh}\varphi (x) {\rm cosh} \varphi (x) + \varphi (x) = x + 
{p \omega \over p_{\perp}^2 + m^2} + {\rm sinh}^{-1}\left( {p \over
\sqrt{p_{\perp}^2 + m^2}}\right)\,,
\end{equation}
for constant electric field. 

This is similar in form to Eqns. (25) and
(26) of \cite{Rau}, the additional sinh $\varphi$ in the numerator
of (\ref{Rausource}) being due to the fact that we have treated
charged scalars rather than fermions in this paper. Thus the form of
the source term plotted in the second halves of Figs. 2-4 is qualitatively similar to those presented in \cite{Rau} by numerical evaluation of an integral similar to (\ref{Rausource}). However, the neglect of the
quantum statistical enhancement (or Pauli blocking) factor
$1 \pm 2 {\cal N}_{\bf k}$ in the integrand of (\ref{nonl}) is valid 
{\it only} in the weak field limit $\lambda \gg 1$. Since that has already been assumed in writing (\ref{Rausource}) one should then properly
evaluate the integral in the same limit. As already remarked
in Section 2 there is no straightforward method of performing
an asymptotic expansion of this integral in real time without losing
the exponentially small Schwinger effect: integrating the cos $(\lambda x)$
term successively by parts will generate the simple adiabatic expansion
which contains no exp $(-\pi \lambda)$ term or step function. 
In the case of weak fields $a^{-1} =\lambda \gg 1$ this effect
is exponentially small in any case, so if one simply
evaluates (\ref{Rausource}) or its equivalent for fermions numerically as in 
Fig. 1 of ref. \cite{Rau} or Fig. 4 of this work, most of the
numerical contribution to what is plotted is contained in the {\it first} (pure $\beta_{\bf k}^{adb}$) term of (\ref{betappr}), which scales like $1/\lambda^3$,
and {\it not} the last term which gives rise to the exponentially small
delta function source of \cite{CasNeu}. Hence multiplication by the factor $\exp (\pi \lambda/2)$ in Eqns. (24) and (25) of ref. \cite{Rau} for weak fields is nugatory, while for strong fields $a^{-1} = \lambda < 1$, the neglect of the factor $1 \pm 2 {\cal N}_{\bf k}$ in (\ref{Rausource}) or Eqn. (25) of \cite{Rau} is not justified.  

\section{Backreaction}
\label{sec:level6}
     
The source term we have derived in (\ref{thesource})
for vacuum initial conditions at $t_0 = -\infty$ must be modified
to include induced creation when there are particles present in
the initial state. Since
\begin{equation}
1 + 2 {\cal N}_{\bf k} = (1 + 2 N_{\bf k}) (1 + 2 |\beta_{\bf k}|^2)
\label{number}
\end{equation}
for $N_{\bf k}$ particles in the initial state, the correct modified
source term in constant electric field is
\begin{equation}
\dot{\cal N}_{\bf k} = S(p, p_{\perp};E) =(1 + 2 N_{\bf k})
S_{vac}(p, p_{\perp};E)\,.
\label{newsourcec}
\end{equation}
In the backreaction problem the electric field will vary with time.
Now the local Markov approximation to the nonlocal Vlasov
equation (\ref{nonl}) consists of using the source term 
(\ref{newsourcec}) with the constant $E$ replaced by $E(t)$ and 
the constant $N_{\bf k}$ by ${\cal N}_{\bf k}(t)$ at the 
{\it local} time of interest. The replacement of $E$ by $E(t)$
is justified if the electric field is slowly varying (the 
quasistationary approximation), while the
replacement of $N_{\bf k}$ by ${\cal N}_{\bf k}(t)$ is justified
if the electric field is weak ($\lambda \gg 1$), since from 
(\ref{number}) the difference between the two is proportional
to $|\beta_{\bf k}|^2$ which is of order $\lambda^{-2}$ and
higher order than the terms we have retained in the asymptotic
expansion of the source term. In this way the statistical
factor $1 \pm 2 {\cal N}_{\bf k}$ has effectively been
removed from the nonlocal integral kernel (\ref{nonl}) and
we have obtained the final form of the source term for use
in backreaction.

Converting the independent variables of the number distribution 
function from canonical momenta $\bf k$ and $t$ to kinetic momenta 
$\bf p$ and $t$ by the definition,
\begin{equation}
{\cal N}_{\bf k}(t)  \equiv  {\cal N}({\bf p} = {\bf k} - e{\bf A}; t)\,,
\end{equation}
we obtain the local Vlasov equation,
\begin{equation}
{\partial\over \partial t} {\cal N}(p, p_{\perp}; t) + e E(t)
{\partial\over \partial p}{\cal N}(p, p_{\perp}; t) = S(p, p_{\perp};E) =
\left(1 + 2{\cal N}(p, p_{\perp}; t)\right) S_{vac}(p, p_{\perp};E)\,, 
\label{newkin}
\end{equation}
for spatially homogeneous fields. Spatial dependence in the distribution
function could be included on the left side of (\ref{newkin}) in the
standard manner, provided it is also slowly varying in space compared to
$c\tau_{cl}$. Together with the Maxwell equation,
\begin{equation}
\ddot A (t) = 2e \int\, [d{\bf p}]\ {p\over\omega } 
{\cal N}(p, p_{\perp}; t)
+2e \int\, [d{\bf p}]\ \omega  
\left(1 + 2{\cal N}(p, p_{\perp}; t)\right) S_{vac}(p, p_{\perp};E)
\label{maxkin}
\end{equation}
(\ref{thesource}), and the defining relations (\ref{rescaled}), (\ref{defxiwz})
and (\ref{defz}) this constitutes the local kinetic approximation to the 
mean field equations.

In order to understand the time scale associated with the
variation of the electric field and therefore the validity
of the quasistationary approximation to the source term by 
that for a constant electric field consider first the
Vlasov-Maxwell system ignoring particle creation.
With the source term set to zero, (\ref{newkin}) can be solved in
closed form, {\it viz.},
\begin{equation}
{\cal N}(p, p_{\perp}; t) = {\cal N}(p+eA(t), p_{\perp}; 0)\,.
\end{equation}
Substituting this solution into (\ref{maxkin}) and linearizing
in $A(t)$ gives
\begin{equation}
\delta \ddot A(t) - 2e^2 \delta A (t)\int\, [d{\bf p}]\, {p\over \omega}\, {\partial\over \partial p}\, {\cal N}(p, p_{\perp}; 0) = 0\,.
\end{equation} 
Integrating the latter expression by parts demonstrates that
the potential (and therefore also the electric field) will
oscillate with a frequency,
\begin{equation}
\omega^2_{pl} = 2e^2 \int\, [d{\bf p}] \, {\cal N}(p, p_{\perp}; 0)
\,{\partial \over \partial p}\left(p\over \omega (p, p_{\perp})\right)\,,
\label{plafreq}
\end{equation}
which is the relativistic plasma frequency. In the nonrelativistic
limit $\omega (p, p_{\perp})$ can be replaced by $m$ and the
integral, $2\int\, [d{\bf p}]\,{\cal N}(p, p_{\perp}; 0) = n$ simply gives the total number density of particles present in the initial state at $t=0$. Then we recover the familiar expression $\omega^2_{pl} \rightarrow e^2 n/m$ 
for the classical plasma oscillation frequency.

This classical plasma frequency may be obtained as well from a linear
response analysis of the quantum mean field equations as follows.
We perturb the vacuum solution for the mode functions,
\begin{equation}
\bar f_{\bf k}(t) = \sqrt{\hbar\over 2\bar\omega_{\bf k}}e^{-i\bar\omega_{\bf k}t}
\,, \qquad \bar\omega_{\bf k} = \sqrt {{\bf k}^2 + m^2}
\end{equation}
with zero electric field by writing
\begin{equation}
f_{\bf k}(t) = \bar f_{\bf k}(t) + \delta f_{\bf k}(t)
\end{equation}
and expand the equations of motion to first order in $\delta f_{\bf k}$,
$A$ and $\dot A$. The linearized mode equation,
\begin{equation}
\left[ {d^2\over dt^2} + \bar\omega_{\bf k}^2\right] \delta f_{\bf k}
= 2ke\,A\, \bar f_{\bf k}
\end{equation}
can be solved by making use of the free retarded Green's function,
\begin{equation}
G_R (t-t'; {\bf k}) = {\sin [\bar\omega_{\bf k}(t-t')] \over \bar\omega_{\bf k}}
\theta (t-t')\,,
\end{equation}
in the form,
\begin{equation}
\delta f_{\bf k}(t) = 2ek\int_0^t\,dt'\,G_R(t-t';{\bf k}) A(t')\bar f_{\bf k}(t')
+ A_{\bf k} \bar f_{\bf k}(t) + B_{\bf k} \bar f_{\bf k}^*(t)\,,
\end{equation}
where $A_{\bf k}$ and $B_{\bf k}$ are constants of integration and 
Re $A_{\bf k} = 0$ in order to preserve the Wronskian condition (\ref{Wron})
under the perturbation. The corresponding linearized Maxwell equation is
\begin{eqnarray}
\ddot A &=& e \int [d{\bf k}] \left\{ 4k\,(1 + 2N_{\bf k})\, {\rm Re}(\bar f_{\bf k}^*
\delta f_{\bf k}) -2eA\,{N_{\bf k}\over \bar\omega_{\bf k}}- eA\,{k^2\over \bar\omega_{\bf k}^3}\,\right\}\label{linmax}\\
&=& 2e^2 \int_0^t\,dt'\, A(t') \int [d{\bf k}]\, {k^2\over \bar\omega_{\bf k}^2}\, 
(1 + 2N_{\bf k})\, \sin [2\bar\omega_{\bf k}(t-t')] 
-e^2 A(t)  \int [d{\bf k}] \left({k^2\over \bar\omega_{\bf k}^3} + 
{2N_{\bf k} \over \bar\omega_{\bf k}}\right) + B(t)\,,\nonumber
\end{eqnarray}
where 
\begin{equation}
B(t) \equiv 2e \int [d{\bf k}]\, {k\over \bar\omega_{\bf k}}\,(1 + 2N_{\bf k}) 
\,{\rm Re}(B_{\bf k}e^{-2i\bar\omega_{\bf k}t})
\end{equation}
is given by the initial perturbation away from the vacuum solution.

The most direct method of solving a linear integral equation such as (\ref{linmax}) is to make use of the Laplace transform,
\begin{equation}
\tilde A(s) \equiv \int_0^{\infty} \,dt\,e^{-st} A(t)\,.
\end{equation}
After some regrouping of terms the Laplace transform of (\ref{linmax})
may be put into the form,
\begin{eqnarray}
&&\tilde A(s) \left\{ s^2 \left[ 1 + {e^2\over 4}\int [d{\bf k}]\, {k^2\over \bar\omega_{\bf k}^5}\,(1 + 2N_{\bf k})\right]
+ 2e^2 \int [d{\bf k}]\, N_{\bf k}\,{\partial \over \partial k}\left({k\over \bar\omega_{\bf k}}\right) \right.\nonumber\\
&&\qquad\qquad\qquad \left. - {e^2s^4\over 4}\int [d{\bf k}]\, {k^2\over \bar\omega_{\bf k}^5}\, {(1 + 2N_{\bf k})\over (s^2 + 4\bar\omega_{\bf k}^2)}\right\} = s A(0) + \dot A(0) +\tilde B(s) \,,
\label{lapmax}
\end{eqnarray}  
where the right hand side depends only upon the initial
data. We notice in (\ref{lapmax}) the presence of the two-particle
threshold at $s^2 = -4\bar\omega_{\bf k}^2$ for the creation of a pair
of charged particles which would give rise to an imaginary part
and damping in the linear Maxwell equation. Since the particles are
massive this imaginary part is zero if we find an oscillatory
solution of the equation with $s= \pm i\omega_{pl}$ and $\omega_{pl} \ll 2m$. 
Such a solution is easily found by setting the expression in curly
brackets in (\ref{lapmax}) to zero and neglecting the $s^4$ term:
\begin{equation}
\omega_{pl}^2 = 2 e_{R, N}^2 \int [d{\bf k}]\, N_{\bf k}{\partial \over \partial k}\left({k\over \bar\omega_{\bf k}}\right)\,,
\label{qpl}
\end{equation}
where
\begin{equation}
{1\over e_{R, N}^2} = {1 \over e^2} + {1\over 4}\int [d{\bf k}]\, {k^2\over \bar\omega_{\bf k}^5}\,(1 + 2N_{\bf k}) \,.
\label{eren}
\end{equation}
In $3+1$ dimensions the combination in (\ref{eren}) is independent of the ultraviolet cut-off and the renormalized value of the charge depends in general on the distribution $N_{\bf k}$. The only requirement on the distribution is that $\omega_{pl}$ in (\ref{qpl}) must be much smaller 
than $2m$, in which limit there is no particle creation
at all and the time independent $N_{\bf k}$ in (\ref{qpl}) may be
identified with the particle density in phase space ${\cal N}({\bf p})$.

Thus the linear response analysis of the quantum mean field theory
gives exactly the same result for the plasma frequency, provided
that the classical charge $e$ appearing in (\ref{plafreq})
is identified with the renormalized charge of the quantum theory
according to (\ref{eren}). This provides 
a consistency check with the classical Vlasov transport description of the plasma, valid in the adiabatic or infrared limit of slowly
varying mean fields, and identifies the proper correspondence limit of
the classical coupling with that in the underlying quantum description.

It is $\omega_{pl}$ that sets the time scale of the variation of the 
electric field in the backreaction problem, {\it i.e.} $\tau_{pl} \simeq
2\pi/\omega_{pl}$. Hence the local, quaistationary approximation to the source term in (\ref{newkin}) requires
that the three time scales obey
\begin{equation}
\omega_{pl}\tau_{qu} \ll  \omega_{pl}\tau_{cl} \ll 1 \,.
\label{ineq}
\end{equation}
If one starts the evolution with zero initial particles then
the distribution function ${\cal N}$ changes from its classical value 
${\cal N}(p+eA(t), p_{\perp}; 0)$ due to the particle creation effect
embodied in the source term. Since the second inequality
requires $\tau_{cl}/\tau_{qu} \sim \lambda \gg 1$
and the source term is exponentially small in $\lambda$,
the number of created particles $n$ and therefore the plasma
oscillation frequency will also be exponentially small in
$\lambda$. Hence the time for enough particles to be produced
to significantly influence the electric field will be exponentially
long and the second inequality in (\ref{ineq}) will also be 
satisfied automatically. Thus our local, quasistationary
approximation scheme for the source term is valid {\it a posteriori},
and we would expect even the cruder approximation of the source
term by (\ref{betappr}) to be not far from correct. 
 
Indeed in previous work we have shown that solving a Vlasov system with
a phenomenological source term of the form 
\begin{eqnarray}
(1+2 {\cal N}(p,p_{\perp}; t)) \vert eE\vert \ln (1 + e^{-\pi\lambda})
\,\delta(p) \, ,
\label{oldsource}
\end{eqnarray}
reproduces results qualititatively similar to the mean field
theory calculation of charged matter field coupled
to a classical electric field \cite{us2}.
In the present work we have shown that no logarithm should be
present in the source term, {\it i.e.} $\ln (1 + e^{-\pi\lambda})$
in (\ref{oldsource}) should be replaced by simply $e^{-\pi\lambda}$, and that in fact, the particle creation event
is continuous and can only be crudely approximated
as a sharp step function in (\ref{betappr}) with some
loss of information about the true time structure of the event,
as demonstrated in Figs. 2-4.
However if the second inequality in (\ref{ineq}) is valid then
the evolution of the mean electric field on the time scale
of $\omega_{pl}^{-1}$ should be affected but little by the further approximation of the source term by a delta function. 

In order to test the validity of this approximation we present
numerical results for the Maxwell-Vlasov system of equations 
(\ref{newkin}) and (\ref{maxkin}) with both the new source term
(\ref{thesource}) and the old delta function source term (but with no logarithm), and compare the results to the exact solution of the mean field evolution of the mode functions (\ref{modeq}) coupled to the Maxwell equation,
(\ref{max}) for scalar QED. The electric field evolution is plotted in Fig. 6
for the three cases. 
We observe that the corrected delta function source 
term gives qualitatively correct results, but the new
source term (\ref{newsourcec}) does a better overall job of tracking 
the mean field evolution, particularly by getting a more accurate value 
of the plasma frequency at late times, where the old source term begins to
drift out of phase. The new source term (\ref{newsourcec}) also drifts
out of phase eventually, but at a much slower rate, or in other words
it more accurately estimates the plasma frequency of the collective
motion. We deliberately chose
moderately large values of the coupling $e=1$ and the initial
electric field $E=1$ in order to amplify the small discrepancy between
the mean field evoluition and that of the new source term. For
evolutions at $e=0.1$ such as in earlier work \cite{us2}, the
discrepancy is negligible on the scale of the plot.
  

\epsfxsize=16cm
\epsfysize=12cm
\vspace{3cm}
\centerline{\epsfbox{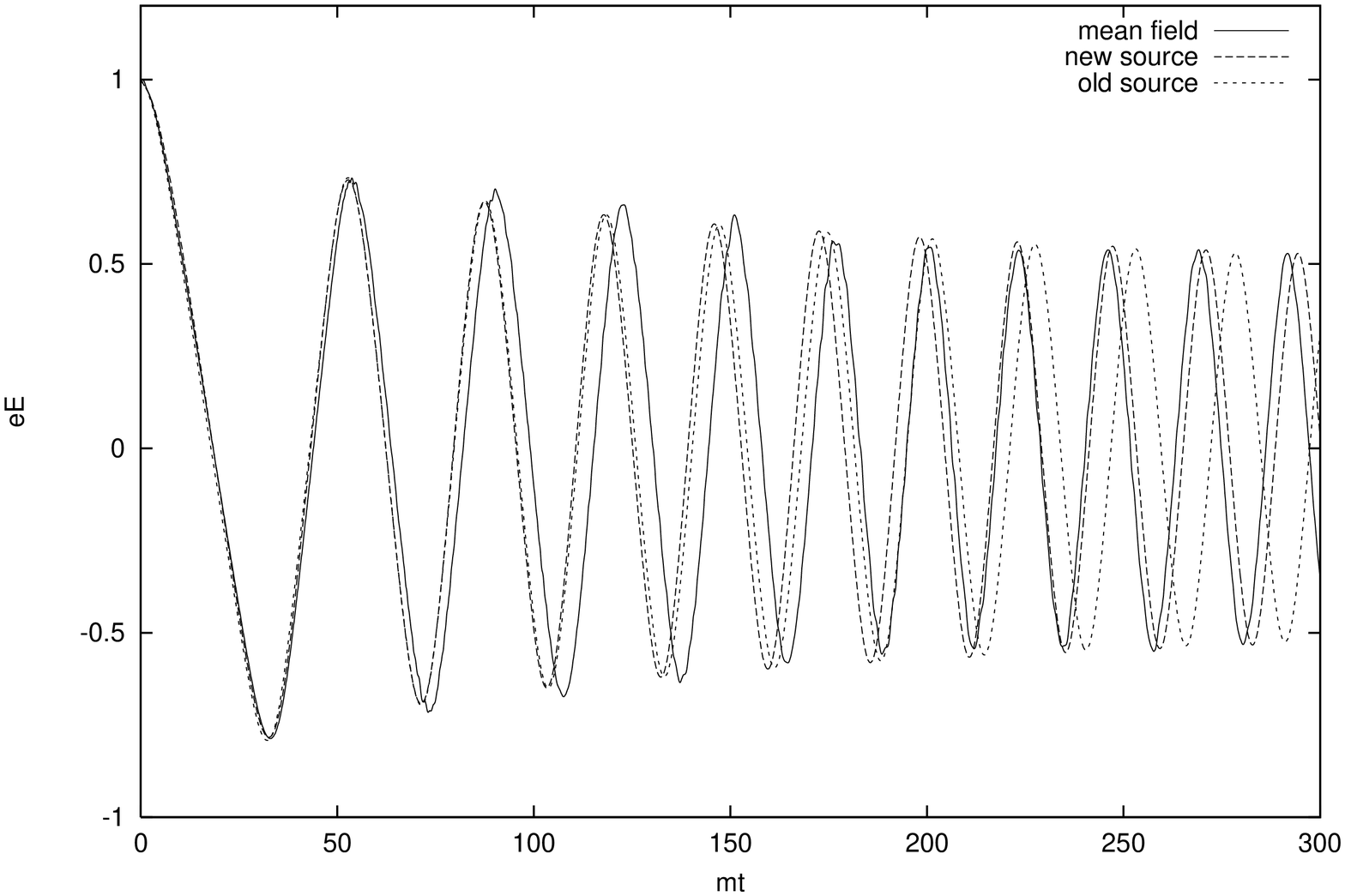}}
{FIG. 6. \small{The evolution of the electric field in one space
dimension, according to
the exact mean field Eqns. (\ref{max}), the new source term (\ref{newsourcec}) derived in this paper, and the old source term used previously
(\ref{oldsource}), but with no logarithm, for initial electric field $eE = e^2 = m^2 =1$, and no particles present in the initial state. The new source term tracks the mean field evolution more accurately than (\ref{oldsource}) which gives a too small plasma oscillation frequency at late times.}}


\epsfxsize=16cm
\epsfysize=12cm
\vspace{3cm}
\centerline{\epsfbox{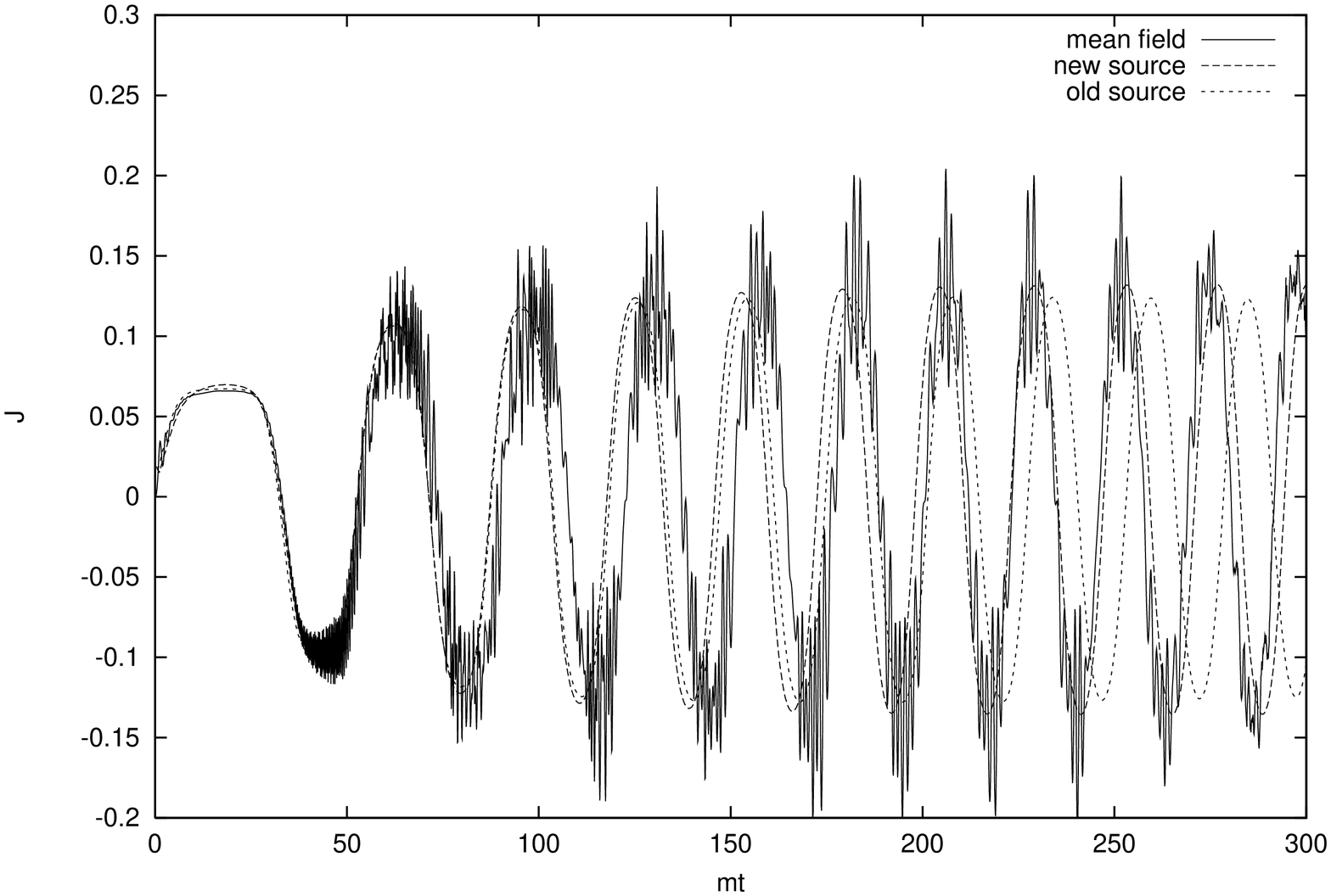}}
{FIG. 7. \small{The evolution of the current in one space dimension, for
the same initial conditions as Fig. 6. Both the source terms for the Vlasov equation neglect the oscillations of the current on the time scale $\tau_{qu} \sim 1$.}}
\vspace{1cm}

In Fig. 7 we display the electric current (\ref{curr3})
for the three evolutions. The new feature observed here are
the rapid oscillations of the quantum mean field evolution on the
time scale $\tau_{qu}$ and their complete absence from the
evolutions with the two local Vlasov source terms which
follow the value of the current averaged over this rapid time scale. 
This is in accord with our previous discussion of the neglect of
such quantum coherence effects in any local transport description.
The particle distribution function ${\cal N}$ is plotted as a function
of $k$ for the mean field and Vlasov evolutions at a particular value of $t$ in Fig. 8. We observe the same quantum coherence effects in the 
mean field evolution here in the rapid oscillations of ${\cal N}$ in 
momentum space on a scale $\Delta k \sim 1/c\tau_{qu}$ as one 
observes as a function of time. Again these oscillations in the
particle distribution are absent in the Vlasov evolutions.
We note also the slightly negative value of the distribution
function in the case of the new Airy source term. This is a
transient effect due to our setting $A_{\bf k}(t_0)$ and $B_{\bf k}(t_0)$
to one and zero respectively. With the more accurate source term
computed from (\ref{betAB}) which takes account of the initial
conditions this artificial negative region is much smaller.
It also grows less and less pronounced as time progresses, and may
be eliminated entirely by binning the distribution in momentum bins.
Some small discrepancy of this kind is to be expected in any
truncation of the unitary field theory evolution by a local
Vlasov source term, unless that source term is always and everywhere
positive, corresponding to a strictly monotonic increase
of total particle number and entropy, according to (\ref{entropy}).
It may be regarded as a rough estimate of
the systematic error induced by the Markov approximation in the
source term (\ref{newsourcec}).

In these numerical evolutions the renormalized charge of the mean
field theory was chosen to be $e_R = 1$ in order to compare
to the Vlasov evolution with unit classical charge $e=1$, according
to (\ref{qpl}) and (\ref{eren}). In $1+1$ dimensions where the simulations 
were performed the charge renormalization is finite and in the vacuum
is given by
\begin{equation}
{m^2\over e_{R, N=0}^2} = {m^2\over e^2} + {1\over 12\pi}\,.
\end{equation}
so that the finite renormalization effects for $e=m=1$ are of 
order $1/12\pi \simeq 0.026$ or a few percent in the range
of the simulations shown in the figures. In the extreme weak coupling
limit ${m^2\over e_R^2} \gg 1$ where the Vlasov approximation
becomes more and more accurate, this finite renormalization effect is completely negligible.  


\vspace{1cm}
\epsfxsize=14cm
\epsfysize=10cm
\centerline{\hspace{-1cm}\epsfbox{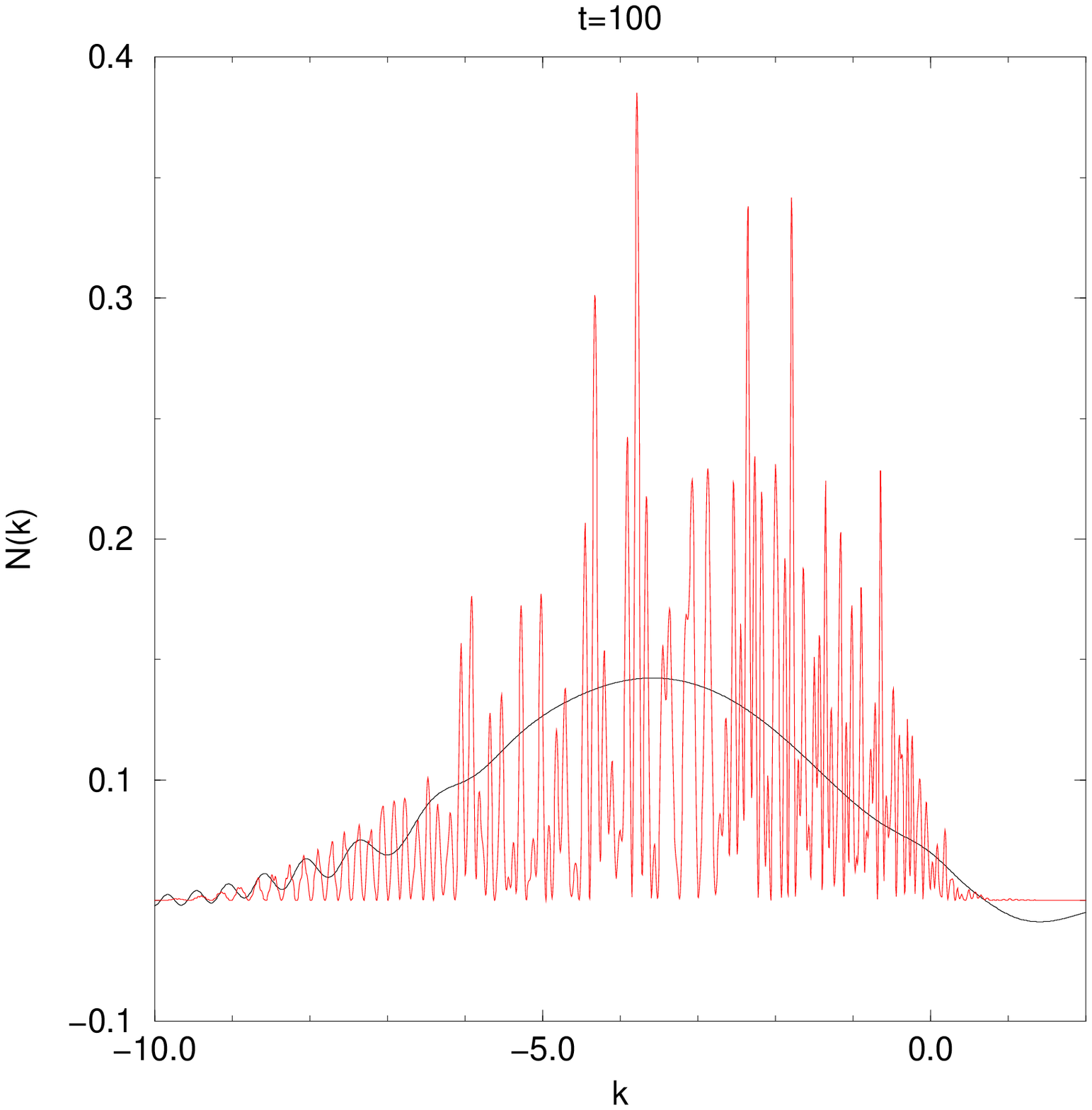}}
\vspace{1cm}
{FIG. 8. \small{The particle distribution $\cal N$ as a function
of canonical momentum $k$ at a fixed time $t=100$ for the mean field
evolution (jagged curve) and the new source term (smooth curve) 
derived in this paper, for the same initial conditions as Figs. 6
and 7. The Vlasov equation with the new source term is approximately
a smooth average of the actual mean field evolution on spatial
scales of order $c\tau_{qu} \sim 1$. The slight dip into negative
values of the smooth curve becomes less and less prominent at later times.}} 
 
\section{Summary and Outlook}

Based on the Hamiltonian description of mean field theory
and the existence of an adiabatic invariant of this evolution,
we identified the (lowest order) adiabatic particle number as the most 
suitable analog for the single particle distribution function of
semiclassical transport theory. Although not unique, this definition 
of particle number involves the fewest number of derivatives 
(namely zero) of the frequency $\omega_{\bf k}$, and hence its
time rate of change is most appropriate for identification as the
source term for the Boltzmann-Vlasov equation, which is first
order in time derivatives. Confirming this identification,
the electric current in this basis has an intuitively appealing and 
simple quasiclassical form, (\ref{curr3}). Since $\dot{\cal N}_{\bf k}$
is already adiabatic order two with this definition of particle
number, including higher order adiabatic corrections in ${\cal N}_{\bf k}$
would be inconsistent with the use of the source term in backreaction
as well, since Maxwell's equations are second order in time.

Analyzing the time dependence of the mean particle number in a constant electric field we derived the rate of pair creation of charged 
scalar particles, and clarified the time scales involved in
the particle production phenomenon. Although formally
equivalent to the quantum Vlasov equation (\ref{nonl}) and
consistent with the general projection method of
Zwanzig applied to the density matrix in the adiabatic
number basis, our approach bypasses the mathematical difficulties 
inherent in the nonlocal integral representation, and does not 
require explicit use of the projection formalism. 
Unlike any direct formal manipulation
of the nonlocal form (\ref{nonl}) or simple WKB expansions, we used a
uniform asymptotic expansion for the local source term,
which retains the Schwinger creation effect at the {\it lowest} order of the expansion. This local source term is not obtained by putting to zero
the phase correlations in the pair creation process, but rather
by the assumption that the actual correlations in a time varying
field can be replaced by those present in a constant field at
at $t =-\infty$. This can only be approximately valid when the 
electric field is very slowly varying in time, so that any
actual phase correlations in the initial state are no longer
important. 

Given the hierarchy of time scales (\ref{ineq}) we showed that
a simple modification of the usual expansion
in terms of exponential functions is nearly adequate for 
most analysis of the collective plasma effects in scalar QED.
The asymptotic expansion in terms of the elementary exponential functions
modified by the step function, which leads to
the ansatz (\ref{betappr}), demonstrates in a simple way the
origin of the linear growth in time in the current which
makes backreaction essential at late times, for {\it any} nonzero
coupling no matter how weak. It also shows
why taking the pair production source term to be proportional to 
$\delta(p)$ was a reasonably good proposal after all, although the use of
the logarithm in the source term of \cite{GleMat,BiaCzy} and subsequent references seems to have been due to a confusion between the rate of particle creation and the vacuum persistence probability. Using this ansatz in conjunction with (\ref{number}) explains the origin of the Bose enhancement source term, which was incorporated in the phenomenological source term (\ref{oldsource}) for physical reasons. The source term obtained in
explicit form from the mode functions in constant electric field (\ref{newsourcec}) is in better agreement with the mean field evolution than the phenomenological source term (\ref{oldsource}), even for
quite large electric fields, although the difference between the
two is not dramatic. 

The methods employed in this paper can be readily extended to other
situations of interest, such as fermions, or the creation of massive particles
by strong gravitational fields in an early universe context. The limit
in which such processes can be described by a semiclassical source
term in a transport approach should be clear from the present work:
one requires a clean separation of the three time scales
$\tau_{qu}$ associated with quantum phase oscillations, $\tau_{cl}$
associated with one particle creation amplitudes,
and $\tau_{pl}$ associated with the collective motion of the
mean field(s). Conversely, it should also be clear that when
such a clean separation does not exist the methods of this
paper cannot be applied, and very likely, no semiclassical
transport approach is appropriate or possible. Unfortunately, 
this includes the cases of most interest in QCD, relativistic heavy-ion 
physics and early universe cosmology, where light or strictly
massless degrees of freedom play an important role. 
If $m=0$ then the low momentum modes will never
behave like classical particles admitting a Blotzmann-Vlasov
description. Even pions are light enough to cause
the hierarchy of time or momentum scales in (\ref{ineq}) to
break down in heavy-ion collisions. In the formation of
disoriented chiral condensates the infrared instability of the
low momentum modes and growth of a large condensate
field by coherence effects is precisely the point. In cases such as
these where Bose condensation plays a central role, the frequencies
$\omega_{\bf k}$ become small or even imaginary, the turning
point(s) of the adiabatic particle number approach or reach
the real time axis, $\tau_{qu}$ becomes large, and no simple quasuclassical 
particle interpretation within the Boltzmann framework is possible. Complementary coherent classical field methods can be developed in this 
regime, matched to a transport description of the hard modes on a case by 
case basis, but only the full field theoretic approach is powerful enough to
encompass all the various cases in a comprehensive fashion.

\vspace{2cm}

{\bf Acknowlegement}
Just as this work was being completed one of the authors (J. M. E.) 
fell seriously ill and passed away a few weeks later. Y. K. and E. M. 
wish to express their profound sadness at this premature loss of
a dear teacher, colleague and friend, and to dedicate this work to his 
memory.

We wish to thank the Institute for Nuclear Theory at the
University of Washington for its hospitality where this work was 
inititiated, and the U.S.~Department of Energy for partial support.
J. M. E.~also wished to thank Professor Walter Greiner and the Institute
for Theoretical Physics at the University of Frankfurt for their kind
hospitality, and to acknowledge support from the Deutsche
Forschungsgemeinschaft and the Ne'eman Chair in Theoretical Nuclear
Physics at Tel Aviv University. Y. K. and E. M. gratefully acknowledge
several enlightening discussions with Salman Habib and Alex Kovner.

\vskip 1 pc
\newpage

\appendix
\section{Density Matrix in the Adiabatic Particle Basis}

In this Appendix we derive the form of the Gaussian density
matrix (\ref{gaussd}) in the adiabatic number basis. Since 
in the case of a spatially homogeneous mean electric field the
density matrix is a product of Gaussians for each wave number $\bf k$,
we consider a single wave number and drop the subscript $\bf k$ in the 
derivation in order to simplify the notation of this Appendix. 

For each wave number we have positively and negatively charged modes
obeying the time dependent harmonic oscillator equation (\ref{modeq}).
Because of this and using Eqns. (\ref{fcooradb}) and (\ref{fmomadb}) the
adiabatic particle basis is that which diagonalizes the Hamiltonian
of the two-dimensional harmonic oscillator,
\begin{equation}
H_{osc} = {1\over 2}\left(\pi^{\dagger}\pi + \omega^2 \varphi^{\dagger}\varphi
+ h. c.\right)
= {\omega\over 2}\left( \tilde a^{\dagger}\tilde a + \tilde a \tilde a^{\dagger}
+ \tilde b^{\dagger} \tilde b + \tilde b \tilde b^{\dagger}\right)
\end{equation}
in the complex representation. The states which diagonalize this 
Hamiltonian are labelled by two quantum numbers $n_+$ and $n_-$ with
energy $\omega (n_+ + n_- + 1)$. In real coordinates,
\begin{equation}
\varphi = {1\over \sqrt 2} (\varphi_1 + i \varphi_2) \equiv {1\over \sqrt 2}r e^{i\theta}
\end{equation}
we can label the states by the radial quantum number $n= n_+ + n_-$
and the angular quantum number $m = n_+ - n_-$ corresponding to the
eigenmodes of the two-dimensional harmonic oscillator,
\begin{equation}
\left( -{1\over 2r}{\partial\over \partial r}r{\partial\over \partial r} -
{1\over 2 r^2} {\partial^2\over \partial \theta^2} +
{1\over 2}\omega^2 r^2 \right) \langle r\, \theta\vert n\,m\rangle = 
\omega (n + 1) \langle r\, \theta\vert n\,m\rangle
\label{schr}
\end{equation}
in polar coordinates. As is well known these wavefunctions are given in
terms of the associated Laguerre polynomials $L^{\alpha}_{\nu}(x)$ in
the form,
\begin{equation}
\langle r\, \theta\vert n\,m\rangle = \left({\omega\over \pi}\right)^{1\over 2}
e^{im\theta} e^{-\omega r^2/2} (\sqrt\omega r)^m 
\left[{\left({n-m\over 2}\right)! \over \left({n+m\over 2}\right)!}\right]^{1\over 2} L^m_{{n-m\over 2}}(\omega r^2)\,.
\label{lagu}
\end{equation}
The normalized eigenstates themselves may be written in the form,
\begin{equation}
\vert n\, m\rangle = { (\tilde a^{\dagger})^{{n+m\over 2}} 
(\tilde b^{\dagger})^{{n-m\over 2}}\over \left[\left({n+m\over 2}\right)!
\left({n-m\over 2}\right)!\right]^{1\over 2}}\vert 0\rangle
\end{equation}
with $m$ taking on the values $-n + 2k$, $k = 0, 1, \dots , n$, so
that $n \pm m$ is an even integer.

In order to transform the density matrix from the coordinate basis
to the adibatic number basis it is easiest first to define the coherent
states,
\begin{eqnarray}
\vert s\, \chi\rangle &\equiv & \exp\left( i\tilde a^{\dagger}s e^{-i\chi}
-i \tilde b^{\dagger} s e^{i\chi} \right)\vert 0\rangle\nonumber\\
&=& \sum_{n=0}^{\infty} {\sum_{m=-n}^n}^{\prime} {s^n e^{-im\chi}
\over \left[\left({n+m\over 2}\right)!
\left({n-m\over 2}\right)!\right]^{1\over 2}}\vert n\, m\rangle\,,
\label{coh}
\end{eqnarray}
where the prime on the sum over $m$ denotes that $m$ is incremented
by even integers. Upon subsituting the explicit wave functions (\ref{lagu})
we find the wave function of these coherent states can be expressed
in the form,
\begin{equation}
\langle r\,\theta\vert s\, \chi\rangle = \left({\omega\over \pi}\right)^{1\over 2} e^{-\omega r^2/2} \sum_{n=0}^{\infty} {\sum_{m=-n}^n}^{\prime} s^n
e^{im(\theta -\chi)} {(\sqrt\omega r)^m 
\over \left({n+m\over 2}\right)!} L^m_{n-m\over 2}(\omega r^2)\,.
\end{equation}
The sums in this expression may be performed in closed form by first
switching the orders of the $n$ and $m=-n +2k$ sums, and making use
of the summation formula \cite{GraRyz},
\begin{equation}
\sum_{n=k}^{\infty} z^n L^{-n + 2k}_{n-k}(x) = z^k \sum_{n=0}^{\infty} z^n
L^{k-n}_n(x) = z^k (1 + z)^k e^{-xz}\,.
\end{equation}
The remaining sum over $k$ from $0$ to infinity is then a pure exponential
and easily performed with the result,
\begin{equation}
\langle r\,\theta\vert s\, \chi\rangle = 
\left({\omega\over \pi}\right)^{1\over 2} \exp\left\{ -{\omega \over 2}r^2
+ 2i rs \sqrt\omega \cos (\chi-\theta) + s^2\right\}\,,
\label{cohwfn}
\end{equation}
or in two-component vector notation,
\begin{equation}
\langle \vec r\, \vert \vec s\,\rangle = \left({\omega\over \pi}\right)^{1\over 2}
\exp\left\{ -{\omega\over 2} \vec r^{\,2} + 2i\sqrt\omega\, \vec r\cdot\vec s + \vec s^{\,2}\right\}\,.
\label{cohvec}
\end{equation}
This $U(1)$ invariant exponential form may be verified also
as the solution of the differential equation,
\begin{equation}
\langle r\,\theta\vert H_{osc}\vert s\, \chi\rangle =
\left( -{1\over 2r}{\partial\over \partial r}r{\partial\over \partial r} -
-{1\over 2 r^2} {\partial^2\over \partial \theta^2} +
{1\over 2}\omega^2 r^2 \right)\langle r\,\theta\vert s\, \chi\rangle
= \omega \left( s {\partial\over \partial s} + 1\right)
\langle r\,\theta\vert s\, \chi\rangle
\end{equation}
obeying the initial condition,
\begin{equation}
\langle r\,\theta\vert s=0\rangle
= \left({\omega\over \pi}\right)^{1\over 2} e^{-{\omega\over 2}r^2}\,
\end{equation} 
which follows from the Schr\"odinger equation (\ref{schr}) and
the definition of the coherent states (\ref{coh}).

The utility of the coherent state basis is apparent from the simple
exponential form of (\ref{cohwfn}) or (\ref{cohvec}), since the
transformation of the density matrix from the original coordinate basis,
\begin{equation}
\langle \vec r^{\ \prime}\vert\rho\vert\vec r\,\rangle =
{1\over 2\pi \xi^2} \exp\left\{ -{(\sigma^2 + 1)\over 8\xi^2} (\vec r^{\,\prime 2}
+ \vec r^{\, 2}) + {i\eta \over 2 \xi} (\vec r^{\,\prime 2} - \vec r^{\,2})
+ {(\sigma^2 - 1)\over 4\xi^2}\vec r^{\,\prime}\cdot\vec r\right\}
\end{equation}  
to the coherent state basis becomes a straightforward exercise in the integration of a product of Gaussians, {\it viz.},
\begin{eqnarray}
\langle \vec s^{\ \prime}\vert\rho\vert\vec s\,\rangle &=& \int d^2\vec r^{\,\prime} \int d^2\vec r\,
\langle \vec s^{\,\prime}\vert\vec r^{\,\prime}\rangle\ \langle \vec r^{\,\prime}\vert\rho\vert\vec r\,\rangle\  \langle \vec r\,\vert\vec s\,\rangle\nonumber\\
&=& {2\omega\xi^2\over B} \exp \left\{ {A\over B}e^{-i\vartheta}\vec s^{\,2} + 
{A\over B}e^{i\vartheta} \vec s^{\,\prime 2} + {C \over B}\vec s \cdot
\vec s^{\,\prime}\right\}
\label{cohden}
\end{eqnarray}
where the real coefficients $A, B, C,$ and $\vartheta$ are given by
\begin{eqnarray}
A\cos \vartheta &=& -\omega^2 \xi^4 + \eta^2\xi^2 + {\sigma^2\over 4}\nonumber\\
A\sin \vartheta &=& -2\omega \eta \xi^2\nonumber\\
B&=& \omega^2\xi^4 + {(\sigma^2 + 1)\over 2}\omega\xi^2 + \eta^2\xi^2 +
{\sigma^2\over 4}\nonumber\\
C&=& (\sigma^2 -1)\omega\xi^2\,.
\label{ABC}
\end{eqnarray}
With this result in hand all that remains to be done is to expand
the coherent state density matrix (\ref{cohden}) in powers of
$s$ and $s'$ to identify the matrix elements of $\rho$ in the
adiabatic particle number basis via
\begin{equation}
\langle s'\,\chi'\vert\rho\vert s\,\chi\rangle =
\sum_{n',n=0}^{\infty} {\sum_m}' {s^{\prime n'}\, s^n e^{im (\chi'-\chi)}
\over\left[\left({n'+m\over 2}\right)!
\left({n'-m\over 2}\right)!\left({n+m\over 2}\right)!
\left({n-m\over 2}\right)!\right]^{1\over 2}}
\ \langle n'\, m\vert\rho\vert n\, m\rangle\,.
\label{rhoexp}
\end{equation}  
The fact that the coherent state density matrix is a function of
only $\chi'-\chi$ and hence only $m' = m$ matrix elements of
$\rho$ appear in the sum is a result of the $U(1)$ invariance
of the density matrix. We also note that in the pure state case,
$\sigma = 1$, $C=0$ and the last dot product cross term in (\ref{cohden})
vanishes, and with it all dependence on $\chi'-\chi$. In that case
only $m=0$ and even $n$ and $n'$ appear in the expansion, and
hence the only nonvanishing matrix elements of the density
matrix are between uncharged states with $n_+ = n_-$. Conversely,
if $\sigma > 1$ this is no longer the case and $\rho$ has
nonvanishing matrix elements also with charged particle states with
$m \neq 0$.  

We first expand the exponential of the dot product,
\begin{eqnarray}
\exp\left({C \over B}\vec s \cdot \vec s^{\,\prime}\right) &=&
\sum_{m=-\infty}^{\infty} i^m J_m \left(-i{C\over B}ss'\right)e^{im(\chi'-\chi)}
\nonumber\\
&=& \sum_{m=-\infty}^{\infty} i^m e^{im(\chi'-\chi)}\sum_{p=0}^{\infty}
{(-)^p\over p! \Gamma (p + m + 1)}\left(-i{C\over 2B}ss'\right)^{m+2p}\,.
\label{bes}
\end{eqnarray}
Multiplying this by the expansion of the exponentials of $s^2$,
\begin{equation}
\exp \left( {A\over B}e^{-i\vartheta}s^2\right) =
\sum_{l =0}^{\infty} {1\over \Gamma(l+1)}\left({A\over B}\right)^l e^{-il\vartheta}s^{2l}\,,   
\end{equation}
and likewise for $s^{\prime 2}$ yields a fourfold sum over $l, l', m$
and $p$. Collecting the powers of $s$ and $s'$ by defining new
summation variables, $n \equiv 2l + m + 2p$ and $n'\equiv 2l' + m + 2p$
we observe that $l + p = {n-m\over 2} \ge 0$ and $l' + p = {n'-m\over 2} \ge 0$
so that $m \le n$ and $ m \le n'$. Also from the presence of the
$\Gamma$ function in the denominator of (\ref{bes}) we observe
that $p + m$ must be nonnegative, which implies ${n'+m\over 2} \ge 0$
and ${n'+m\over 2} \ge 0$. Hence $m \ge -n$ and $m \ge -n'$ as well,
and we can write
\begin{eqnarray}
&&\exp\left\{  {A\over B}e^{-i\vartheta}\vec s^{\,2}+ 
+{A\over B}e^{i\vartheta} \vec s^{\,\prime 2} + {C \over B}\vec s \cdot
\vec s^{\,\prime}\right\} =\nonumber\\
&& \sum_{n,n'=0}^{\infty}\sum_{m=-M}^{M\,\prime}  e^{i\vartheta (n'-n)/2} e^{im(\chi - \chi')} s^{\prime n'} s^n  
\sum_{p=0}^{\infty}{i^m(-)^p\over p! \Gamma (p + m +1)} {\left( -{iC\over 2B}\right)^{m + 2p}\over \Gamma ({n-m\over 2} -p +1)}{\left({A\over B}\right)^{{n+n'\over 2} -m -2p} \over \Gamma ({n'-m\over 2} -p +1)}\,,
\label{expand}
\end{eqnarray}
where $M = min (n,n')$. We also note that ${n \pm m\over 2}$ and ${n' \pm m\over 2}$ are necessarily integers in this expression.

Because of the $\Gamma$ functions in the denominator the final sum over $p$
in (\ref{expand}) terminates at $p = min (n,n')$. However it is convenient 
to retain the formal infinite range of $p$ and make use of the relation 
for the $\Gamma$ function,
\begin{equation}
{1 \over \Gamma (1-z)} = \Gamma (z) {\sin (\pi z)\over \pi}
\end{equation}
for $z = p -{n-m \over 2}, p -{n'-m \over 2}, -{n-m \over 2}$ and 
$-{n'-m \over 2}$, temporarily continuing ${n-m \over 2}$ and ${n'-m \over 2}$
to noninteger values to avoid the appearance of divergences in the
intermediate steps. In this way the sum over $p$ is recognized
as the expansion for the hypergeometric function $_2F_1 \equiv F$,
\begin{eqnarray}
&&\sum_{p=0}^{\infty}{i^m(-)^p\over p! \Gamma (p + m +1)} {\left( -{iC\over 2B}\right)^{m + 2p}\over \Gamma ({n-m\over 2} -p +1)}{\left({A\over B}\right)^{ -m -2p} \over \Gamma ({n'-m\over 2} -p +1)} \nonumber\\
&&=\sum_{p=0}^{\infty}
\left({C\over 2A}\right)^{m + 2p}{\Gamma (p -{n-m\over 2})
\Gamma (p -{n'-m\over 2})\Gamma(m+1)\over \Gamma (-{n-m\over 2})
\Gamma (-{n'-m\over 2}) p! \Gamma(p+m+1)}{1\over m!\left({n-m\over 2}\right)!
\left({n'-m\over 2}\right)!} \nonumber\\
&& =\left({C\over 2A}\right)^m {1\over m!\left({n-m\over 2}\right)!
\left({n'-m\over 2}\right)!} F\left({m-n\over 2}, {m-n'\over 2}; m+1; {C^2\over
4A^2}\right)\,.
\end{eqnarray}
and we secure
\begin{eqnarray}
&&\langle s'\,\chi'\vert\rho\vert s\,\chi\rangle = {2\omega\xi^2 \over B}
\times \label{rexpan}\\
&& \sum_{n',n=0}^{\infty} {\sum_{m=-M}^M}' {s^{\prime n'}\, s^n e^{im (\chi'-\chi)} \over m!\left({n-m\over 2}\right)! \left({n'-m\over 2}\right)!}
\left({A\over B}\right)^{n+n'\over 2} \left({C\over 2A}\right)^m
e^{i\vartheta (n'-n)/2} F\left({m-n\over 2}, {m-n'\over 2}; m+1; {C^2\over
4A^2}\right)\,.\nonumber
\end{eqnarray}
The finite sum represented by the hypergeometric function with
integral indices may be expressed in terms of Jacobi polynomials
$P^{(\alpha, \beta)}_{\nu}$ if desired, through the relation \cite{Erd}
\begin{equation}
P^{({n'-n\over 2}, m)}_{n-m\over 2}\left({1+z\over 1-z}\right)
= {\left({n+m\over 2}\right)!\over m! \left({n-m\over 2}\right)!}
\left({z\over 1-z}\right)^{n-m\over 2}F\left({m-n\over 2}, {m-n'\over 2}; m+1; z\right)\,.
\end{equation}

Comparing (\ref{rexpan}) to the general form (\ref{rhoexp}) 
we may identify the matrix elements of the density matrix in the adiabatic particle basis to be
\begin{eqnarray}
&&\langle n'\, m\vert\rho\vert n\, m\rangle =\label{rhoparta}\\
&& {2\omega\xi^2 \over B}
\left({A\over B}\right)^{n+n'\over 2}\left({C\over 2A}\right)^m
{e^{i\vartheta (n'-n)/2} \over m!}\left[{\left({n + m\over 2}\right)!
\left({n' + m\over 2}\right)!\over \left({n-m\over 2}\right)!
\left({n'-m\over 2}\right)!}\right]^{1\over 2}
F\left({m-n\over 2}, {m-n'\over 2}; m+1; {C^2\over 4A^2}\right)\,.\nonumber
\end{eqnarray}
This is the desired result. It may be expressed in terms of the magnitude
and phase of the Bogoliubov transformation from the Heisenberg
basis to the time dependent adiabatic particle basis introduced
in the text. In fact, making use of the definitions (\ref{xietadef}),
(\ref{Bog}) and (\ref{thetdef}) we have
\begin{eqnarray}
2\omega\xi^2 &=& \sigma ({\rm cosh} 2\gamma - {\rm sinh} 2 \gamma \cos 
\vartheta)\qquad {\rm and}\nonumber\\
2\xi\eta &=& -\sigma {\rm sinh} 2 \gamma \sin \vartheta\,,
\end{eqnarray}
and from (\ref{ABC}) we obtain
\begin{eqnarray}
A &=& \omega \xi^2 \sigma {\rm sinh} 2\gamma\qquad {\rm and}\nonumber\\
B &=& 2\omega \xi^2 \left[\sigma {\rm cosh}^2 \gamma + 
\left({\sigma - 1\over 2}\right)^2\right]\,,
\end{eqnarray}
so that finally,
\begin{eqnarray}
&&\langle n'\, m\vert\rho\vert n\, m\rangle ={(\sigma {\rm sinh} \gamma {\rm cosh} \gamma)^{{n+n'\over 2}-m}
\over \left[\sigma {\rm cosh}^2 \gamma + \left({\sigma - 1\over 2}\right)^2\right]^{{n+n'\over 2} + 1}} \left({\sigma^2 -1\over 4\sigma}\right)^m e^{i\vartheta (n'-n)/2}\times\nonumber\\
&&\qquad\qquad{1\over m!}\left[{\left({n + m\over 2}\right)!
\left({n' + m\over 2}\right)!\over \left({n-m\over 2}\right)!
\left({n'-m\over 2}\right)!}\right]^{1\over 2}
F\left({m-n\over 2}, {m-n'\over 2}; m+1; {(\sigma^2-1)\over 
4\sigma^2 {\rm sinh}^2 2\gamma}\right)\,.
\label{rhopart}
\end{eqnarray}
Since $\rho$ is symmetric under charge conjugation we have
\begin{equation}
\langle n',\, -m\vert\rho\vert n,\, -m\rangle = \langle n'\, m\vert\rho\vert n\, m\rangle\,,
\end{equation}
which implies that the mean charge $\sum_m m \langle n'\, m\vert\rho\vert n\, m\rangle = 0$. The fact that $\rho$ has nonvanishing matrix elements with states of nonzero $m$ implies that the fluctuations of the charge about 
its mean value is nonzero in the general case of $\sigma \neq 1$. Otherwise 
the most important feature of the general result (\ref{rhopart}) for the
density matrix in the adiabatic particle basis for the purposes of the
discussion in the text is that all the off diagonal
matrix elements for $n\neq n'$ are rapidly varying functions of time
because of the appearance of the phase $\vartheta$. Since all the
phase correlation information of the function $\cal C$ of (\ref{pcor})
resides in these off diagonal elements, while the average
adiabatic particle number $\cal N$ is sensitive only to the diagonal
matrix elements of $\rho$, the Markov limit of quantum Vlasov
equation corresponds to replacement of the density matrix by
only its diagonal matrix elements in this basis.  

In the pure state case $\sigma =1$, $F =1$, and the only nonvanishing
matrix elements of $\rho$ have $m=0$ and $n= 2\ell, n'=2\ell'$
both even. The general result (\ref{rhopart}) simplies considerably
in this case to
\begin{equation}
\langle 2\ell'\, m=0\vert\rho\vert 2\ell\, m=0\rangle\bigg\vert_{_{\sigma =1}} = 
{\rm sech}^2 \gamma ({\rm tanh} \gamma)^{\ell + \ell'} e^{i\vartheta (\ell'-\ell)}\,,
\label{finalrhod}
\end{equation}
which yields the result (\ref{rhodiag}) quoted in the text. We should note
that this expression differs from that used in previous work 
\cite{us3},
since in the present derivation the distinguishability of the positive
and negative charged particles was taken into account, leading
to a two-dimensional harmonic oscillator problem with a $U(1)$
invariance, while the expression Eqn. (15) of the second of refs. \cite{us3} 
or Eqn. (5.24) and the entire Appendix of the third of refs. \cite{us3}
was based on a single scalar particle species. This is appropriate
for the real uncharged $\Phi^4$ theory considered in the last of \cite{us3}, whereas (\ref{finalrhod}) is the correct expression for the charged particle
case. 
 
\end{document}